\documentclass[10pt,a4paper]{article}%{elsart}%amsmath,amssymb]%{revtex4}
\usepackage{amssymb,amsmath} 
 \usepackage{amsthm}  
\usepackage{graphicx}
\usepackage{fullpage} 
\usepackage{color}
\usepackage{bm}
\usepackage[all]{xy} 
\usepackage{makeidx}
\makeindex
\begin{document} 
%%%%%%%%%%%%%%%%
\newtheorem{Def}{Definition}[section]
\newtheorem{Thm}{Theorem}[section]
\newtheorem{Proposition}{Proposition}[section]
\newtheorem{Lemma}{Lemma}[section]
\theoremstyle{definition}
\newtheorem*{Proof}{Proof}%[section]
\newtheorem{Example}{Example}[section]
\newtheorem{Postulate}{Postulate}[section]
\newtheorem{Corollary}{Corollary}[section]
\theoremstyle{remark}
\newtheorem{Remark}{Remark}[section]
\newcommand{\beq}{\begin{equation}}
\newcommand{\beqa}{\begin{eqnarray}}
\newcommand{\eeq}{\end{equation}}
\newcommand{\eeqa}{\end{eqnarray}}
\newcommand{\non}{\nonumber}
\newcommand{\fr}[1]{(\ref{#1})}
\newcommand{\abs}{\mathrm{abs}}
\newcommand{\B}{\mathrm{B}}
\newcommand{\G}{\mathrm{G}}
\newcommand{\can}{\mathrm{can}}
\newcommand{\Diff}{\mbox{Diff}}
\newcommand{\Id}{\mathrm{Id}}
\newcommand{\bC}{\mbox{\boldmath {$C$}}}
\newcommand{\bK}{\mbox{\boldmath {$K$}}}
\newcommand{\bp}{\mbox{\boldmath {$p$}}}
\newcommand{\bx}{\mbox{\boldmath {$x$}}}
\newcommand{\by}{\mbox{\boldmath {$y$}}}
\newcommand{\bz}{\mbox{\boldmath {$z$}}}
\newcommand{\bF}{\mbox{\boldmath {$F$}}}
\newcommand{\bJ}{\mbox{\boldmath {$J$}}}
\newcommand{\bT}{\mbox{\boldmath {$T$}}}
\newcommand{\bR}{\mbox{\boldmath {$R$}}}
\newcommand{\bDelta}{\mathbf{\Delta}}
\newcommand{\ve}{{\varepsilon}}
\newcommand{\e}{\mathrm{e}}
\newcommand{\dr}{\mathrm{d}}
\newcommand{\Dr}{\mathrm{D}}
\newcommand{\F}{\mathrm{F}}
\newcommand{\mbbE}{\mathbb{E}}
\newcommand{\mbbP}{\mathbb{P}}
\newcommand{\tA}{\widetilde A}
\newcommand{\tB}{\widetilde B}
\newcommand{\tC}{\widetilde C}
\newcommand{\uD}{\underline{\Delta}}
\newcommand{\chQ}{\check{Q}}
\newcommand{\chN}{\check{N}}
\newcommand{\chS}{\check{S}}
\newcommand{\cheta}{\check{\eta}}
\newcommand{\chX}{\check{X}}
\newcommand{\chY}{\check{Y}}
\newcommand{\cA}{{\cal A}}
\newcommand{\cB}{{\cal B}}
\newcommand{\cC}{{\cal C}}
\newcommand{\cD}{{\cal D}}
\newcommand{\cE}{{\cal E}}
\newcommand{\cF}{{\cal F}}
\newcommand{\cG}{{\cal G}}
\newcommand{\cH}{{\cal H}}
\newcommand{\cI}{{\cal I}}
\newcommand{\cK}{{\cal K}}
\newcommand{\cL}{{\cal L}}
\newcommand{\cM}{{\cal M}}
\newcommand{\cN}{{\cal N}}
\newcommand{\cO}{{\cal O}}
\newcommand{\cP}{{\cal P}}
\newcommand{\cR}{{\cal R}}
\newcommand{\cS}{{\cal S}}
\newcommand{\cT}{{\cal T}}
\newcommand{\cU}{{\cal U}}
\newcommand{\cV}{{\cal V}}
\newcommand{\cY}{{\cal Y}}
\newcommand{\tcA}{\widetilde{\cal A}}
\newcommand{\cAone}{   \underset{(1)}{ {\cal A} }     }
\newcommand\wh[1]{\widehat{#1}}
\newcommand{\DD}{{\cal D}}
\newcommand{\sympro}{\overset{s}{\otimes}}
\newcommand{\hash}{\#}
\newcommand{\GamCLamM}[1]{{\Gamma\mathbb{C}\Lambda^{{#1}}\cal{M}}}
\newcommand{\GamLamA}[1]{{\Gamma\Lambda^{{#1}}\cal{A}}}
\newcommand{\GamLamB}[1]{{\Gamma\Lambda^{{#1}}\cal{B}}}
\newcommand{\GamLamC}[1]{{\Gamma\Lambda^{{#1}}\cal{C}}}
\newcommand{\GamLamS}[1]{{\Gamma\Lambda^{{#1}}\cal{S}}}
\newcommand{\GamLamM}[1]{{\Gamma\Lambda^{{#1}}\cal{M}}}
\newcommand{\GamLamN}[1]{{\Gamma\Lambda^{{#1}}\cal{N}}}
\newcommand{\GTM}{{\Gamma T\cal{M}}}
\newcommand{\GTB}{{\Gamma T\cal{B}}}
\newcommand{\GTC}{{\Gamma T\cal{C}}}
\newcommand{\GT}[1]{{\Gamma T {#1}}}
\newcommand{\normM}[2]{\left(  {#1}\, , \, {#2} \right)}
\newcommand{\normU}[2]{\left\{ {#1}\, , \, {#2} \right\}}
\newcommand{\diag}[1]{\mbox{diag}\{\, {#1}\,\}}
\newcommand{\GtC}[2]{\Gamma T^{#1}_{#2}{\cal C}}
\newcommand{\GtM}[2]{\Gamma T^{#1}_{#2}{\cal M}}
\newcommand{\GtN}[2]{\Gamma T^{#1}_{#2}{\cal N}}
\newcommand{\Gt}[3]{\Gamma T^{#1}_{#2}{#3}}
\newcommand{\equp}[1]{\overset{\mathrm{#1}}{=}}
\newcommand{\lequp}[1]{\overset{\mathrm{#1}}{\leq}}
\newcommand{\gequp}[1]{\overset{\mathrm{#1}}{\geq}}
\newcommand{\lup}[1]{\overset{\mathrm{#1}}{<}}
\newcommand{\gup}[1]{\overset{\mathrm{#1}}{>}}
\newcommand{\rightup}[1]{\overset{\mathrm{#1}}{\longrightarrow}}
\newcommand{\leftup}[1]{\overset{\mathrm{#1}}{\Longleftarrow}}
\newcommand{\Leftrightup}[1]{\overset{\mathrm{#1}}{\Longleftrightarrow}}
\newcommand{\leftrightup}[1]{\overset{\mathrm{#1}}{\longleftrightarrow}}
\newcommand{\nLongleftup}{\Longleftarrow\hspace{-4.5mm}\diagup}
\newcommand{\fun}[1]{{#1}^{\hash}}
\newcommand{\wt}[1]{\widetilde{#1}}
\newcommand{\ol}[1]{\overline{#1}}
\newcommand{\sff}[1]{ {\sf{#1}}}
\newcommand{\sfL}{ {\sf L}}
\newcommand{\ddiv}{\mbox{div}}
\newcommand{\rank}{\mbox{rank}\,}
\newcommand{\bkt}[2]{\left\{\,{#1}\, ,\,{#2}\,\right\}}
\newcommand{\inp}[2]{\left\langle\,{#1}\, ,\,{#2}\,\right\rangle}
\newcommand{\inpr}[2]{\left(\,{#1}\, ,\,{#2}\,\right)}
\newcommand{\ave}[1]{\left\langle\,{#1}\, \right\rangle}
\newcommand{\avgg}[2]{\left\langle\,{#1}\, \right\rangle_{#2}}
\newcommand{\Leg}{{\mathfrak{L}}}
\newcommand{\ic}{\sqrt{-1}\,}
\newcommand{\ii}{\imath}
\newcommand{\bracket}[3]{\left[\,{#1},{#2}\,\right]_{#3}}
\newcommand{\LM}[2]{{\Lambda^{{#1}}_{{#2}}\,\cal{M}}}
\newcommand{\TM}[2]{{T^{{#1}}_{{#2}}\,\cal{M}}}
%%%%%%%%%%%

%\renewcommand{\labelenumi}{\Roman{enumi}}
\renewcommand{\labelenumi}{{\bf\arabic{enumi}. }}
% 
% To be black
%\newcommand{\void}[1]{}
%\renewcommand{\color}{\void}

%\renewcommand{}
%%%%%%%%%%
 
%\def\man{{M}}
%\def\GamLamM#1{{\Gamma \Lambda^{#1}\cal{M}}}
%\begin{footnotesize} 

%%%%%%%%%%%%%%%%%%%%%%%%%%%%%%%%%%%%%%%%%%%%%%%%%%%%%%
\title{Legendre submanifolds in contact manifolds as attractors and geometric nonequilibrium thermodynamics 
}

%%%%%%%%%%%%%%%%%%%%%%%%%%%%%%%%%%%%%%%%%%%%%%%%%%%%%%
%\author{First Author${}^\dag$ and Second Author${}^\ddag$}
\author{Shin-itiro Goto\footnote{{\tt sgoto at ims.ac.jp}}\\
Institute for  Molecular Science, \\
38 Nishigo-Naka, Myodaiji, Okazaki 444-8585, Japan
}

\date{28 July 2015}%June 10, 2015}%\today}%\quad 
%The date of resubmission to the revised version will be added}
\maketitle
%%%%%%%%%%%%%%%%%
\begin{abstract}%
%%%%%%%%%%%%%%%%%
It has been proposed that equilibrium thermodynamics is described on 
Legendre submanifolds in contact geometry.  
It is shown in this paper that Legendre submanifolds 
embedded in a contact manifold can be expressed 
as attractors in phase space for a certain class of contact Hamiltonian 
vector fields. 
By giving a physical interpretation that 
points outside the Legendre submanifold can represent 
nonequilibrium states
of thermodynamic variables,   
in addition to that points of a given 
Legendre submanifold can represent equilibrium states of the variables, 
this class of contact Hamiltonian vector fields  
is physically interpreted as a class of relaxation processes, 
in which thermodynamic variables achieve 
an equilibrium state from a nonequilibrium state through a time evolution,
a typical nonequilibrium phenomenon. 
Geometric properties of such vector fields on 
contact manifolds are characterized after introducing 
a metric tensor field on a contact manifold. 
It is also shown that 
a contact manifold and a strictly convex function induce a 
lower dimensional dually flat space used in information geometry 
where a geometrization of equilibrium statistical mechanics is constructed.
Legendre duality on contact manifolds is explicitly stated throughout. 

%%%%%%%%%%%%%%%%
\end{abstract}
%%%%%%%%%%%%%
\section{Introduction}
%%%%%%%%%%%%%%%%%%%
There have been several attempts to geometrically describe 
equilibrium and nonequilibrium thermodynamics,
and the most standard one may be based on contact geometry that is 
an odd-dimensional counterpart of symplectic geometry.  
Hermann is one of the first to formulate geometric equilibrium thermodynamics
based on contact geometry \cite{Hermann1973},
and his work has shown that the so-called Legendre submanifolds 
embedded in a contact manifold are suitable 
for describing the first law of thermodynamics and Legendre transforms of  
equilibrium thermodynamics. 
Although there are some outcomes for equilibrium systems 
along with this context after Hermann's 
book, there are still many questions that are needed to be resolved. 
It should be noted that there are other approaches to geometrically 
describe equilibrium 
thermodynamics. These include the work of Weinhold \cite{Weinhold} 
and that of Ruppeiner. In both of them, Hessian matrices 
of functions  
are used for expressing components of metric tensor fields with respect to 
particular coordinates. Thus, they are related to Hessian geometry.
These works have been followed by a number of papers, and   
related findings were summarized in Ref.\cite{Ruppeiner1995}. 
In these days, they are applied to black hole thermodynamics 
(See Refs.\cite{Aman2007} and \cite{Bravetti-2014PRD}, for example).    
In addition how to geometrically 
connect an equilibrium distribution function of  
microscopic variables and equilibrium thermodynamics has been known 
\cite{Brody2009}.  
This offers the use of information geometry, 
a geometrization of mathematical statistics \cite{AN}, to study equilibrium 
statistical mechanics and equilibrium thermodynamics. 

Mrugala {\it et al.}\cite{Mrugala1990} suggested a way to combine    
geometric equilibrium thermodynamics
and equilibrium statistical mechanics 
in which probability distributions play a role. 
Here and in what follows 
geometric equilibrium and nonequilibrium thermodynamics 
are identified with the ones developed with contact geometry, rooted in  
Hermann's idea.    
In the work of Mrugala {\it et al.}, 
an equation being equivalent to the first law 
of thermodynamics is used as a constraint 
for placing equilibrium states in a contact manifold.   
One observes that a part of their formulation 
is similar to that of information geometry. 
In information geometry emphasis is placed on Legendre duality,  
and therefore, it is well matched to equilibrium thermodynamics. 
We then feel that such Legendre duality 
discussed at equilibrium states may be promoted to or survive at some 
nonequilibrium states.  
Furthermore, we expect that there are some connections among 
Hessian geometry, information geometry, 
and thermodynamics, since there is an approach to study information 
geometry with Hessian geometry \cite{Matsuzoe2013}. 
In accordance with the work of Mrugala {\it et al.}, 
there are some extensions of geometric thermodynamics. These include 
Ref.\cite{Bravetti-Sep-2014} in which above mentioned Legendre duality    
and relations between thermodynamics and  
a contact Riemannian manifold 
were elaborated, where a contact Riemannian manifold  
consists of a contact manifold  
with additional data including a metric tensor field.

A class of nonequilibrium thermodynamics may also be described by the use of 
contact geometry. 
However, there is little consensus in the literature on how best to 
give physical interpretations of general points of a contact manifold.  
To study a time-dependent nonequilibrium phenomenon,   
one needs to introduce a dynamical system  and its phase space. 
An appropriate dynamical system may be the so-called 
contact Hamiltonian system, and the phase space for this 
 may be a contact manifold.
The physical interpretation of such dynamical systems varies.   
For example Jurkowki assumed in his paper \cite{Jurkowski2000} that 
the contact Hamiltonian system with his particular contact Hamiltonian  
gives deformations of submanifolds of thermodynamic equilibrium states. 
In Ref.\cite{Bravetti-Sep-2014}, a particular form of 
contact Hamiltonian was proposed, and their contact Hamiltonian flow is 
interpreted as a near-equilibrium process. 
It should be noted that 
there are other directions to develop the use of contact 
Hamiltonian systems. These include Refs.\cite{Schaft2007} and 
\cite{Bravetti-Dec-2014}.

In this paper,  
relations among contact manifolds, Hessian manifolds, and 
mathematical objects in information geometry 
including statistical manifolds and 
dually flat spaces are elucidated. Then,   
we shall adopt the view that 
points outside Legendre submanifolds of 
a contact manifold 
can express a class of nonequilibrium states, where  
a Legendre submanifold is given.
Our physical interpretation of points outside Legendre submanifolds are  
nonequilibrium states of thermodynamic variables, which is  
the same as that of Ref.\cite{Bravetti-Sep-2014}.
In particular, we propose   
a class of contact Hamiltonian systems that can physically be interpreted as 
a class of  relaxation processes. Here, a relaxation process is 
that thermodynamic variables achieve an equilibrium state 
from a nonequilibrium one through time evolution, 
one of typical nonequilibrium phenomena.  
As an application of this general theory 
a nonequilibrium spin system that exhibits a relaxation process is shown.  
The system derived in the framework of contact geometry is shown 
to include one derived with a master equation for such a spin system. 
In geometric language,  
a relaxation process is an integral curve that connects  
a point outside a Legendre submanifold of a contact manifold  
and a point of a given Legendre submanifold.   
In general, 
when a metric tensor field and a 
connection are introduced on a contact manifold, one can 
geometrically characterize a 
dynamical system. We investigate our class of contact 
Hamiltonian systems from this view point. 
Throughout this paper, emphasis is placed on Legendre duality  
inside and outside Legendre submanifolds.  

To illustrate some of the issues above this paper is organized as follows. 
In \S\ref{sec-review}, tools and ideas developed in contact geometry 
are summarized. These are necessary to state our claims. In addition, 
physical interpretations of mathematical tools and ideas are postulated,  
some existing results are also summarized. 
In \S\ref{sec-statistical-manifold-legendre-submanifold}, 
theorems are given for providing relations 
among various objects in information geometry 
and Legendre submanifolds of a contact manifold.
In \S\ref{sec-attractor},
explicit forms of contact Hamiltonians for describing relaxation processes
and an example are given.    
In \S\ref{sec-calculations-with-metric},
with a metric tensor field various quantities 
involving the relaxation processes
and the so-called quasi-stationary processes are calculated for characterizing
the nonequilibrium processes. 
Finally \S\ref{sec-summary} summarizes 
our paper and discusses some of future works.

%%%%%%%%%%%%%%%%%%%%%%%%%%%%%%%%%
\section{Contact manifold and physical quantities}
\label{sec-review}
%%%%%%%%%%%%%%%%%%%%%%%%%%%%%%%%%%
\subsection{Mathematical preliminaries}
\label{sec-Mathematical-preliminaries}
%%%%%%%%%%%%%%%%%%%%%%%%%%%%%%%%%%%%%%%%%%%%%%%%%%%%
In this subsection, we give a brief summary of contact geometry in order to 
describe theorems that will be shown in the following sections. 
Throughout this paper, geometric objects are assumed smooth, 
a set of vector fields on a manifold $\cM$ is denoted $\GTM$,
the tangent space at $\xi\in\cM$ as $T_{\xi}\cM$, 
a set of $q$-forms $\GamLamM{q}$ with $q\in\{0,\ldots,\dim\cM\}$,  
and a set of tensor fields $\GtM{q'}{q}$ with 
$q,q'\in\{0,1,\ldots\}$. 
To express tensor fields, the direct product is denoted $\otimes$.  
Einstein notation, when an index variables appear twice in a single 
term it implies summation of all the values of the index, 
is adopted. The exterior derivative acting on $\GamLamM{q}$ is denoted 
$\dr:\GamLamM{q}\to\GamLamM{q+1}$, and 
the interior product operator with $X\in\GTM$ as $\ii_X:\GamLamM{q}\to\GamLamM{q-1}$.  
Given a map $\Phi$ between two manifolds, the pull-back is denoted $\Phi^*$, 
and the push-forward $\Phi_*$.  
Then, one can define the Lie derivative acting on tensor fields 
with respect to 
$X\in\GTM$ denoted $\cL_{X}:\GtM{q'}{q}\to\GtM{q'}{q}$. 
It follows that $\cL_{X}\beta=(\ii_X\dr+\dr\ii_X)\beta,$ 
for any $\beta\in\GamLamM{q}$, which is referred to as the Cartan formula. 
%%%%%%%%%%%%
\begin{Def}
%%%%%%%%%%
(Contact manifold) : 
Let $\cC$ be a $(2n+1)$-dimensional manifold, and $\lambda$ a 
one-form on $\cC$ such that
$$
\lambda\wedge\underbrace{\dr \lambda\wedge\dr\lambda\cdots\wedge\dr\lambda}_{n}
\neq 0,
$$
at any point on $\cC$.
If $\cC$ carries $\lambda$, then $(\,\cC,\lambda\,)$ 
is referred to as a contact manifold and $\lambda$ a contact form. 
%%%%%%%%%%
\end{Def}
%%%%%%%%%%
%%%%%%%%
\begin{Remark}
%%%%%%%%%%%%
The $(2n+1)$-form $\lambda\wedge\dr\lambda\wedge\cdots\wedge\dr\lambda$ 
can be used for a volume form. 
%%%%%%%%
\end{Remark}
%%%%%%%%

There is a known standard local coordinate system.
%%%%%%%%%%%%%
\begin{Thm}
%%%%%%%%%%%%
(Canonical coordinates) : There exist local $(2n+1)$ coordinates 
$(x,p,z)$ with $x=\{x^1,\ldots,x^n\}$ and $p=\{p_1,\ldots,p_n\}$,  
in which $\lambda$ has the form 
\beq
\lambda
=\dr z-p_a\dr x^a.
\label{contact-form-standard-Darboux}
\eeq
The $(x,p,z)$ are referred to as the canonical coordinates, 
or the Darboux coordinates. 
%%%%%%%%%%
\end{Thm} 
%%%%%%%%%%%
In addition to the above coordinates,
ones in which $\lambda$ has the 
form $\lambda=\dr z+p_a\dr x^a$ are also used in the literature.
In this paper, \fr{contact-form-standard-Darboux} is used. 

Given a contact manifold, 
there exists a unique vector field that is defined as follows.
%%%%%%%
\begin{Def}
%%%%%%
(Reeb vector field) : 
Let $(\cC,\lambda)$ be a contact manifold, and $R$ a vector field on $\cC$. 
If $R$ satisfies 
$$
\ii_R \dr\lambda
=0\qquad\mbox{and}\qquad 
\ii_R\lambda
=1,
$$
then $R\in\GTC$ is referred to as the Reeb vector field, or  
the characteristic vector field.  
%%%%%%%
\end{Def}
%%%%%%

%%%%%%%
\begin{Remark}
%%%%%%%
From the definition of $R$, one has 
\beq
\cL_{R}\lambda
=0,
\label{remark-Reeb-vector-Lie-derivative}
\eeq
where $\cL_{R}$ is the Lie derivative with respect to $R$. 
To show \fr{remark-Reeb-vector-Lie-derivative},  
one uses the Cartan formula.
%%%%%%%
\end{Remark}
%%%%%%%

As mentioned, $R$ is uniquely determined when $\lambda$ is given, 
and a coordinate expression for $R$ is given as follows.
%%%%%%%
\begin{Thm}
%%%%%%
(Coordinate expression of the Reeb vector field) : 
Let $(\cC,\lambda)$ be a contact manifold, and $R$ the Reeb vector field.
If the canonical coordinates $(x,p,z)$ are such that 
$\lambda=\dr z\pm p_a\dr x^a$  
with $x=\{x^1,\ldots,x^n\}$ and $p=\{p_1,\ldots,p_n\}$, then 
$$
R=\frac{\partial }{\partial z}.
$$  
%%%%%%%
\end{Thm}
%%%%%%

To formulate equilibrium thermodynamics geometrically, one needs 
the following definition.
%%%%%%%%%
\begin{Def}
\label{def-Legendre-submanifold}
%%%%%%%
(Legendre submanifold) : 
Let $(\,\cC,\lambda\,)$ be a contact manifold, 
$\cA$  a submanifold of $\cC$, and $\Phi:\cA\to\cC$ an embedding.   
If $\cA$ is 
a maximal dimensional integral submanifold such that $\Phi^*\lambda=0$, 
then $\cA$ is referred to as a Legendre submanifold.
%%%%%%%%
\end{Def}
%%%%%%%%

The following theorem states the dimension of a Legendre submanifold for 
a given contact manifold.
%%%%%%%%%%%%
\begin{Thm}
\label{thm:max-integral-submanifold-is-n}
(Maximal dimensional integral submanifold) : 
Let $(\,\cC,\lambda\,)$ be a $(2n+1)$-dimensional contact manifold, 
$\cA$  a submanifold, and $\Phi:\cA\to\cC$ an embedding.  
The maximal dimensional integral submanifolds such that $\Phi^*\lambda=0$ 
is equal to $n$.  
\end{Thm}
%%%%%%%%%%
%%%%%%%%
\begin{Remark}
%%%%%%%%%
Combining Theorem \ref{thm:max-integral-submanifold-is-n}
 and Definition \ref{def-Legendre-submanifold}, 
one concludes that 
the dimension of any Legendre submanifold of a 
$(2n+1)$-dimensional contact manifold is $n$.
%%%%%%%%
\end{Remark}
%%%%%%%

The following theorem shows
the explicit expressions of Legendre submanifolds in terms of 
canonical coordinates. 
%%%%%%%%%%%
\begin{Thm}
\label{theorem-Legendre-submanifold-theorem-Arnold}
%%%%%%%%%%% 
(Local expression of Legendre submanifold, \cite{Arnold1976}) :  
Let $(\,\cC,\lambda\,)$ be a $(2n+1)$-dimensional contact manifold,  
and $(x,p,z)$ the canonical coordinates such that $\lambda=\dr z-p_a\,\dr x^a$  
with $x=\{x^1,\ldots,x^n\}$ and $p=\{p_1,\ldots,p_n\}$. 
For any partition $I\cup J$ of the set of indices $\{1,\ldots,n\}$ into 
two disjoint subsets $I$ and $J$, and for a function $\phi(x^J,p_I)$ of 
$n$ variables $p_i,i\in I$, and $x^j,j\in J$ the $(n+1)$ equations
\beq
x^i=-\,\frac{\partial\phi}{\partial p_i},\qquad
p_j=\frac{\partial\phi}{\partial x^j},\qquad 
z=\phi-p_i\frac{\partial\phi}{\partial p_i}
\label{Legendre-submanifold-theorem-Arnold}
\eeq
define a Legendre submanifold. Conversely, every 
Legendre submanifold of $(\,\cC,\lambda\,)$ in a neighborhood of any point is    
defined by these equations for at least one of the $2^n$ possible choices 
of the subset $I$.
%%%%%%%%%%%
\end{Thm}
%%%%%%%%%%% 
%%%%%%%%
\begin{Def}
%%%%%%%%
(Legendre submanifold generated by a function) : 
The function $\phi$ used in 
Theorem\,\ref{theorem-Legendre-submanifold-theorem-Arnold}  
is referred to as a generating function of the Legendre submanifold. 
If a  Legendre submanifold $\cA$ is expressed 
as \fr{Legendre-submanifold-theorem-Arnold}, then $\cA$ is referred to as 
a Legendre submanifold generated by $\phi$. 
%%%%%%%%%
\end{Def}
%%%%%%%%%

The following are examples of local expressions for the 
 Legendre submanifolds.  
They will be used in the following sections.
%%%%%%%
\begin{Example}
%%%%%%%
\label{example-Arnold-Legendre-submanifold-psi}
Let $(\cC,\lambda)$ be a $(2n+1)$-dimensional contact manifold,
$(x,p,z)$ the canonical coordinates such that $\lambda=\dr z-p_a\dr x^a$  
with $x=\{x^1,\ldots,x^n\}$ and $p=\{p_1,\ldots,p_n\}$, 
and $\psi$ a function of $x$ only.   
The Legendre submanifold $\cA_{\psi}$ generated by $\psi$ with 
$\Phi_{\cC\cA\psi}:\cA_{\psi}\to\cC$ being an embedding 
is such that 
\beq
\Phi_{\cC\cA\psi}\cA_{\psi}
=\left\{\ (x,p,z)\in\cC \ \bigg|\ 
p_j=\frac{\partial\psi}{\partial x^j},\ \mbox{and}\ 
z=\psi(x),\quad j\in \{1,\ldots,n\}
\ \right\}. 
\label{example-psi-Legendre-submanifold}
\eeq
One can easily verify that $\Phi_{\cC\cA\psi}^{\ \ \ \  *}\lambda=0$. 
Note that the relation between this $\psi$ and $\phi$ of   
\fr{Legendre-submanifold-theorem-Arnold} is $\psi(x)=\phi(x)$ with 
$J=\{1,\ldots,n\}$.
%%%%%%%
\end{Example}
%%%%%%%
%%%%%
\begin{Example}
%%%%%
\label{example-Arnold-Legendre-submanifold-varphi}
Let $(\cC,\lambda)$ be a $(2n+1)$-dimensional contact manifold,
$(x,p,z)$ the canonical coordinates such that $\lambda=\dr z-p_a\dr x^a$ 
with $x=\{x^1,\ldots,x^n\}$ and $p=\{p_1,\ldots,p_n\}$, 
and $\varphi$ a function of $p$ only. 
The Legendre submanifold $\cA_{\varphi}$ generated by $\varphi$ 
with $\Phi_{\cC\cA\varphi}:\cA_{\varphi}\to\cC$ being 
an embedding is such that 
\beq
\Phi_{\cC\cA\varphi}\cA_{\varphi}
=\left\{\ (x,p,z)\in\cC \ \bigg|\ 
x^i=\frac{\partial\varphi}{\partial p_i},\ \mbox{and}\ 
  z=p_i\frac{\partial\varphi}{\partial p_i}-\varphi(p),\quad i\in \{1,\ldots,n\}
\ \right\}. 
\label{example-varphi-Legendre-submanifold}
\eeq
One can easily verify that $\Phi_{\cC\cA\varphi}^{\ \ \ \ *}\lambda=0$.
Note that 
the relation between this $\varphi$ and $\phi$ of   
\fr{Legendre-submanifold-theorem-Arnold} is $\varphi(p)=-\,\phi(p)$ with
$I=\{1,\ldots,n\}$.
%%%%%
\end{Example}
%%%%%

One can choose a function $\psi$ in 
Example\,\ref{example-Arnold-Legendre-submanifold-psi} to generate $\cA_{\psi}$ 
and 
$\varphi$ in Example \ref{example-Arnold-Legendre-submanifold-varphi} to 
$\cA_{\varphi}$  
independently, and in this case, there is no relation between 
$\cA_{\psi}$ and $\cA_{\varphi}$ in general. 
On the other hand, when $\psi$ is strictly convex, and $\varphi$ 
is carefully chosen, it will be shown in the next section that 
there is a relation between $\cA_{\psi}$ and $\cA_{\varphi}$. 
To discuss such a case, 
the following transform should be introduced. The convention is suitably
adopted to that in information geometry. 
Note that several conventions exist in the literature. 
%%%%%%%%
\begin{Def}
%%%%%%%%
(Total Legendre transform) : 
Let $\cM$ be an $n$-dimensional manifold,
$x=\{x^1,\ldots,x^n\}$ coordinates, 
and $\psi$ a function of $x$.
Then, the total Legendre transform of $\psi$ with respect to $x$ 
is defined to be  
\beq
\Leg[\psi](p)
:=\sup_{x}\left[\,x^ap_a-\psi(x)\,\right],
\label{def-total-Legendre-transform}
\eeq
where $p=\{p_1,\ldots,p_n\}$.
%%%%%%%%
\end{Def}
%%%%%%%%

From this definition, one has several formulas that will be used in the 
following sections. 
%%%%%%%%%
\begin{Thm}
%%%%%%%%
\label{theorem-Legendre-tranform-formula}
(Formulas involving the total Legendre transform) : 
Let $\cM$ be an $n$-dimensional manifold, $x=\{x^1,\ldots,x^n\}$ coordinates,
$\psi\in\GamLamM{0}$ a strictly convex function of $x$ only, and 
$\varphi$ the function of $p$ obtained by the total Legendre 
transform of $\psi$ with respect to $x$ where $p=\{p_1,\ldots,p_n\}$ :
$\varphi(p)=\Leg[\psi](p)$. Then, for each $a$ and fixed $p$, the equation 
$$
p_a
=\left.\frac{\partial\psi (x)}{\partial x^a}\right|_{x=x_*}
=\frac{\partial\psi (x_*)}{\partial x_*^a},
$$
has the unique solution
$x_*^a=x_*^a(p), (a\in\{1,\ldots,n\})$. In addition it follows that 
$$
\varphi(p)
=x_*^ap_a-\psi(x_*),\qquad
\frac{\partial\varphi}{\partial p_a}
=x_*^a,\qquad
\delta_{b}^{a}
=\frac{\partial^2\psi}{\partial x_*^b\partial x_*^l}\frac{\partial^2\varphi}{\partial p_a\partial p_l},
$$
and 
$$
\det\left(\frac{\partial^2\,\psi}{\partial x^a\partial x^b}\right)>0,
\qquad
\det\left(\frac{\partial^2\,\varphi}{\partial p_a\partial p_b}\right)>0.
$$
%%%%%%%%%
\end{Thm}
%%%%%%%%

A way to describe dynamics on a contact manifold is to introduce a 
continuous diffeomorphism with a parameter. 
First one defines a diffeomorphism on a contact manifold.
%%%%%%%%%%5
\begin{Def}
\label{def-contact-diffeomorphism}
%%%%%%%% 
(Contact diffeomorphism) : 
Let $(\,\cC,\lambda\,)$ be a $(2n+1)$-dimensional 
contact manifold, 
and $\Phi:\cC\to\cC$  a diffeomorphism. If it follows that 
$$
\Phi^*\lambda
=f\,\lambda,
$$  
where $f\in\GamLamC{0}$ is a function that does not vanish on 
any point of $\cC$, then the map $\Phi$ is referred to as 
a contact diffeomorphism. 
%%%%%%%%%%5
\end{Def}
%%%%%%%% 
%%%%%%%%%
\begin{Remark}
%%%%%%%%
The transformed one-form in Definition \ref{def-contact-diffeomorphism}
is also a contact form since 
$$
f\lambda\wedge\underbrace{\dr (f\lambda)\wedge\cdots\wedge \dr(f\lambda)}_{n}
=f^{n+1}\lambda\wedge\underbrace{\dr \lambda\wedge\cdots\wedge \dr\lambda}_{n}
\neq 0.
$$
%%%%%%%%
\end{Remark}
%%%%%%%%%
%%%%%%%%
\begin{Remark}
%%%%%%%%%
It follows
that $\Phi$ preserves the contact structure, 
$\ker\lambda:=\{\,X\in\GTC\, |\,\ii_X\lambda=0\,\}$, 
but does not preserve the original 
contact form.
%%%%%%%%
\end{Remark}
%%%%%%%%%

In addition to this diffeomorphism, one 
can introduce  one-parameter groups as follows.
%%%%%%%%%%5
\begin{Def}
%%%%%%%% 
(One-parameter group of continuous contact transformations) : 
Let $(\,\cC,\lambda\,)$ be a $(2n+1)$-dimensional contact manifold, 
and $\Phi_{t}:\cC\to\cC$ a diffeomorphism with $t\in\mathbb{R}$ that satisfies
$\Phi_0=\Id_{\,\cC}$ and $\Phi_{t+s}=\Phi_t\circ\Phi_s, (t,s\in\mathbb{R})$ 
where $\Id_{\,\cC}$ is such that $\Id_{\,\cC}\xi=\xi$ for all $\xi\in\cC$. 
If it follows that  
$$
\Phi_{t}^*\lambda
=f_t\,\lambda,
$$  
where $f_t\in\GamLamC{0}$ is a function that does not vanish on 
any point of $\cC$, then the $\Phi_t$ is referred to as 
a one-parameter group of continuous contact transformations. 
If $t,s\in T$ with some $T\subset \mathbb{R}$ then it is referred to as 
a one-parameter local transformation group of continuous transformations.
%%%%%%%%%%5
\end{Def}
%%%%%%%% 
A contact vector field is defined as follows. 
%%%%%%%%%%
\begin{Def}
%%%%%%%%
(Contact vector field) : 
Let $(\cC,\lambda)$ be a contact manifold, and $X$ a vector field on $\cC$.
If $X$ satisfies 
$$
\cL_X\lambda
=f\,\lambda,
$$
where $f$ is non-vanishing function on $\cC$, then 
$X$ is referred to as a contact vector field. 
%%%%%%%%%%
\end{Def}
%%%%%%%%

A  one-parameter (local) transformation groups
is realized by integrating the following vector field.
%%%%%%%%
\begin{Def}
%%%%%%%
(Contact vector field associated to a contact Hamiltonian) : 
Let $(\cC,\lambda)$ be a contact manifold, $h$  a function on $\cC$, and 
$X_h$  a vector field.   
If $X_h\in\GTC$ satisfies 
\beq
\ii_{X_h}\lambda
=h\qquad\mbox{and}\qquad 
\ii_{X_h}\dr \lambda
=-\,(\,\dr h-(Rh)\,\lambda\,),
\label{def-contact-vector-Hamiltonian}
\eeq
then $X_h$ is referred to as a contact vector field \underline{associated} to 
a function $h$ or a contact Hamiltonian vector field. 
In addition $h$ is referred to as a contact Hamiltonian. 
%%%%%%% 
\end{Def}
%%%%%%
Note that one cannot interpret contact Hamiltonian vector fields  
as classical Hamiltonian vector fields on symplectic manifold  in general.  

The definition \fr{def-contact-vector-Hamiltonian} 
and the Cartan formula give 
$\cL_{X_h}\lambda=(Rh)\,\lambda$, from which 
one has the following. 
%%%%%%%
\begin{Thm}
%%%%% 
Let $(\cC,\lambda)$ be a contact manifold, $h$ a contact Hamiltonian,  
and $X_h$ a contact Hamiltonian vector field. If $Rh\in\GamLamC{0}$ 
does not vanish at 
any point on $\cC$,   then $X_h$ is a contact vector field.
%%%%%%%
\end{Thm}
%%%%% 

Local expressions of a contact Hamiltonian vector field
\fr{def-contact-vector-Hamiltonian} are straightforwardly calculated 
as follows.
%%%%%%%
\begin{Thm}
%%%%% 
(Local expression of contact Hamiltonian vector field) : 
Let $(\cC,\lambda)$ be a $(2n+1)$-dimensional contact manifold, 
$h$ a contact Hamiltonian,  
$X_h$ a contact Hamiltonian vector field, and $(x,p,z)$ 
the canonical coordinates 
such that $\lambda=\dr z-p_a\dr x^a$ 
with $x=\{x^1,\ldots,x^n\}$ and $p=\{p_1,\ldots,p_n\}$. Then,   
$$
X_h
=\dot{x}^a\frac{\partial}{\partial x^a}
+\dot{p}_a\frac{\partial}{\partial p_a}
+\dot{z}\frac{\partial}{\partial z},
$$
where $\dot{}$ denotes the differential 
with respect to a parameter $t\in\mathbb{R}$, or $t\in T$ 
with some $T\subset\mathbb{R}$, 
and 
\beq
\dot{x}^a
=-\,\frac{\partial h}{\partial p_a},\qquad
\dot{p}_a
=\frac{\partial h}{\partial x^a}+p_a\frac{\partial h}{\partial z},\qquad
\dot{z}
=h-p_a\frac{\partial h}{\partial p_a},\qquad a\in\{1,\ldots,n\}.
\label{contact-Hamiltonian-vector-components}
\eeq
%%%%%%%
\end{Thm}
%%%%% 
The following theorem is well-known, and has been 
used in the literature of geometric thermodynamics.  
%%%%%%%%%%%%%
\begin{Thm}
%%%%%%%%%%%%
(Tangent vector field of Legendre submanifold realized 
by contact Hamiltonian vector field, \cite{Mrugala1991}) : 
Let $(\cC,\lambda)$ be a contact manifold, $\cA$ a Legendre submanifold, and 
$h$ a contact Hamiltonian. Then,  
the contact Hamiltonian vector field is tangent to $\cA$ if and only if 
$h$ vanishes on $\cA$. 
%%%%%%%%%%%
\end{Thm}
%%%%%%%%%%

%%%%%%%%5%%%%%%%%%%%%%%%%%%%
\subsection{Physical interpretations of mathematical objects  }
%%%%%%%%%%%%%%%%%%%%%%%%%%
In this subsection,  
we give our physical quantities and physical interpretations of mathematical 
objects introduced in the previous subsection. 

Since nonequilibrium thermodynamics is discussed in this paper, 
we need the following postulate and definition.
%%%%%%%%%%%%5
\begin{Postulate}
%%%%%%%%%%%%%%
(Equilibrium thermodynamic state and Legendre submanifold, 
\cite{MrugalaX}) : 
Let $\cA_{\phi}$ be 
the Legendre submanifold  generated by $\phi(x^J,p_I)$ 
in Theorem\,\ref{theorem-Legendre-submanifold-theorem-Arnold}, 
$\Phi:\cA_{\phi}\to\cC$ its embedding.  
Then, points of $\Phi\cA_{\phi}$ express equilibrium states.  
If there is $\Phi'\cA_{\phi^{\prime}}$ such that it is 
diffeomorphic to $\Phi\cA_{\phi}$,   
then we identify $\Phi\cA_{\phi}$ with $\Phi'\cA_{\phi^{\prime}}$. 
%%%%%%%%%%%\%%
\end{Postulate}
%%%%%%%%%%%%%%%%
%%%%%%%%%%%
\begin{Remark}
%%%%%%%%%%%%%
The variables $x^J,p_I$ and the function $\phi$ of $(x^J,p_I)$ in 
Theorem\,\ref{theorem-Legendre-submanifold-theorem-Arnold} 
specify the equilibrium state of a system.
%%%%%%%%%%%%%
\end{Remark}
%%%%%%%%%%%%

%%%%%%%%%%%%%%%%%
\begin{Def}
%%%%%%%%%%%%%%%
(Equilibrium states and nonequilibrium states) : 
If a thermodynamic variable is not at equilibrium, then the state is 
referred to as a nonequilibrium state.
%%%%%%%%%%%%%%%%%
\end{Def}
%%%%%%%%%%%%%%%
We restrict ourselves to a simple case by postulating the following.
%%%%%%%%%%%%%%%
\begin{Postulate}
%%%%%%%%%%%%
(Spatial homogeneity of thermodynamic systems) : 
Every thermodynamic system 
is assumed spatially homogeneous even at nonequilibrium states. 
%%%%%%%%%%%%%%%
\end{Postulate}
%%%%%%%%%%%%
%%%%%%%%%%%
\begin{Remark}
%%%%%%%%%%
Throughout this paper, spatial coordinates will not be introduced.  
%%%%%%%%%
\end{Remark}
%%%%%%%%5

The basic notations used below follow the standard thermodynamics.
%%%%%%%%%%%%%%%%%%%%5
\begin{Def}
%%%%%%%%%%%%%%%%%%%
\label{notation-thermodynamic-variables}
(Thermodynamic variables and physical quantities at equilibrium) :  
The symbol $S$ denotes entropy, 
$V$ volume, $N_k$ the number of moles for species $k$, 
$T_{\abs}$ the absolute temperature, $P$ pressure, 
$\mu_k$ chemical potential for species $k$, 
$U$ internal energy, $\Omega_{\G}$ the grand canonical potential, and $\cF$ 
the Helmholtz free energy. The abbreviations 
$\beta_{\abs}:=(k_{\B}T_{\abs})^{-1}$ with $k_{\B}$ being the Boltzmann constant and 
$\ln Z_{\G}:=-\beta_{\abs}\Omega_{\G}$  will be used. 
The extensive thermodynamic variables are assumed normalized.   
Here, normalized thermodynamic variables 
are obtained by dividing unnormalized thermodynamic variables by amount of 
substance. 
The introduced variables above 
are referred to as thermodynamic variables or physical quantities.
These are defined at equilibrium.  
%%%%%%%%%%%%%%%%%
\end{Def}
%%%%%%%%%%%%%%%

%%%%%%%%%%%
\begin{Postulate}
%%%%%%%%%%%%%%%%%
(Thermodynamic variables at nonequilibrium states) : 
The thermodynamic variables at equilibrium in Definition 
\ref{notation-thermodynamic-variables} can be extended to 
those at some nonequilibrium states. 
%%%%%%%%%%%
\end{Postulate}
%%%%%%%%%%%%%%%%%
%%%%%%%%%%
\begin{Remark}
%%%%%%%%%%
We do not distinguish notationally between thermodynamic variables at 
equilibrium and those at nonequilibrium states. 
%%%%%%%%
\end{Remark}
%%%%%%%%

The following are often used in the literature.  
%%%%%%%%%%%%5
\begin{Postulate}
%%%%%%%%%%%%%%
(Thermodynamic variables and canonical coordinates in contact geometry, \cite{MrugalaX}) : 
Let $(\cC,\lambda)$ be a $(2n+1)$-dimensional contact manifold, 
and $(x,p,z)$ the canonical coordinates 
such that $\lambda=\dr z-p_a\dr x^a$  
with $x=\{x^1,\ldots,x^n\}$ and $p=\{p_1,\ldots,p_n\}$.   
Then, $(x,p,z)$ can physically represent thermodynamic variables. 
In particular, 
$z$ represents either a thermodynamic potential or 
a dimensionless one,   
$(x,p)$ a set of pairs of other normalized extensive and 
intensive variables such that $p_a$ is conjugate to $x^a$ for 
each $a\in\{1,\ldots,n\}$. 
%%%%%%%%%%%\%%
\end{Postulate}
%%%%%%%%%%%

%%%%%%%%%%%%
\begin{Def}
%%%%%%%%%%%5
(Gibbs one-form and thermodynamic phase space) : 
Let $(\cC,\lambda)$ be a contact manifold. 
When $\lambda$ is written in terms of physical quantities, 
$\lambda$ is referred to as the Gibbs one-form, and 
$\cC$ as thermodynamic phase space. 
%%%%%%%%%%%%
\end{Def}
%%%%%%%%%%%5
%%%%%%%%%%
\begin{Remark}
%%%%%%%%%%
Thermodynamic phase space includes equilibrium thermodynamic systems.
%%%%%%%%%%
\end{Remark}
%%%%%%%%%%

The following examples of the Gibbs one-form are well-known. 
%%%%%%%%%%%
\begin{Example}
%%%%%%%%%%%
(Energy representation,\cite{MrugalaX}) : 
Identify $(x,p,z)$  as 
$$
(x^1,x^2,x^3,\ldots;p_1,p_2,p_3,\cdots;z)
\Longleftrightarrow
(S,V,N_1,\ldots;T_{\abs},-P,\mu_1,\ldots;U).
$$   
Then, the Gibbs one-form $\lambda^U$ 
in terms of the introduced variables is 
$$
\lambda^U
:=\dr U-T_{\abs}\dr S+P\dr V-\mu_k\dr N^k,
$$ 
where all the variables in the right hand side are assumed independent.  
The first law holds where $\lambda^U$ vanishes.
%%%%%%%%%%%
\end{Example}
%%%%%%%%%%%
%%%%%%%%%%%
\begin{Example}
%%%%%%%%%%%
(Entropy representation, \cite{MrugalaX}) : 
Identify $(x,p,z)$ as  
$$
(x^1,x^2,x^3,\ldots;p_1,p_2,p_3,\cdots;z)
\Longleftrightarrow
(U,V,N_1,\ldots;1/\,T_{\abs},P/\,T_{\abs},-\mu_1/\,T_{\abs},\ldots;S).
$$   
Then, the Gibbs one-form $\lambda^S$ 
in terms of the introduced variables is 
$$
\lambda^S
:=\dr S-\frac{1}{T}\dr U-\frac{P}{T}\dr V-\frac{\mu_k}{T}\dr N^k,
$$ 
where all the variables in the right hand side are assumed independent.  
The first law holds where $\lambda^S$ vanishes.
%%%%%%%%%%%
\end{Example}
%%%%%%%%%%%

So far  standard interpretations have been given.   
On the other hand, 
the following postulates may not be common in the literature. 

To state postulates, the following definition is needed. 
%%%%%%%%%%
\begin{Def}
%%%%%%%%%%
(Outside Legendre submanifold generated by $\phi$ of a contact manifold) : 
Let $\cA_{\phi}$ be 
the Legendre submanifold  generated by $\phi(x^J,p_I)$ in 
Theorem\,\ref{theorem-Legendre-submanifold-theorem-Arnold},   
and $\Phi:\cA_{\phi}\to\cC$ its embedding. 
Then,  
$$
\cN_{x^J,p_I,\phi}
:=\cC\,\setminus\Phi\cA_{\phi}
$$ 
is referred to as outside Legendre submanifold generated by $\phi$. 
In addition, let $\cA_{\phi'}$ be 
the Legendre submanifold  generated by $\phi'(x^{J'},p_{I'})$ in 
Theorem\,\ref{theorem-Legendre-submanifold-theorem-Arnold},   
and $\Phi':\cA_{\phi'}\to\cC$ its embedding. 
When $\Phi'\cA_{\phi'}$ 
is diffeomorphic to $\Phi\cA_{\phi}$,  
the set $\cN_{[x^J,p_I,\phi]}$ is defined such that $\cN_{x^{J'},p_{I'},\phi'}$ 
is identical to $\cN_{x^J,p_I,\phi}$.
%%%%%%%%%%
\end{Def}
%%%%%%%%%%

%%%%%%%%%%5
\begin{Postulate}
%%%%%%%%%%%%%%
\label{postulate-nonequilibrium-outside-Legendre}
(Nonequilibrium states and outside Legendre submanifold of a contact manifold) : 
The set $\cN_{[x^J,p_I,\phi]}$
physically represents some nonequilibrium states.
%%%%%%%%%%%\%%
\end{Postulate}
%%%%%%%%%%%
%%%%%%%%%
%%%%%%
\begin{Example}
\label{example:spin-equilibrium}
%%%%%%%
(Nonequilibrium states of a spin system and outside Legendre submanifold) : 
Consider one spin system with an 
external constant magnetic field $H$ 
in contact with a heat bath of 
temperature $T_{\abs}$, where the physical dimension of $H$ is an energy. 
Let $m$ be the magnetization, and $z=- \,\cF/\,(\,k_{\B}T_{\abs}\,)$ 
a dimensionless negative 
Helmholtz free energy. Then, if the relations
$$
m=\tanh\frac{H}{k_{\B}T_{\abs}}\quad\mbox{and}\quad 
z=\ln\cosh\frac{H}{k_{\B}T_{\abs}}+\ln 2 
$$
hold for given values of $H$ and $T_{\abs}$, 
then the system is at equilibrium. If not, 
the system is at nonequilibrium. 
%%%%%%
\end{Example}
%%%%%%%
A relaxation dynamics of this particular spin system 
will be studied in 
\S\ref{subsection-spin-system-relaxation}. 
In addition, the constant $\ln 2$ in this example will appear in 
a calculation for the equilibrium state in that subsection.   

%%%%%%%%%%%%%%
\begin{Remark}
%%%%%%%%%%%%%%
This example is generalized as follows. Let $x=\{x^1,\ldots,x^n\}$ be  
thermodynamic variables, $p=\{p_1,\ldots,p_n\}$ 
their conjugate variables,  
$z$ a dimensionless negative Helmholtz free energy of $x$, 
and $\psi$ its equilibrium value. 
Then, if 
$p_a=\partial\psi/\partial x^a$ and $z=\psi$ hold, 
then the system is at equilibrium. 
This is consistent with \fr{example-psi-Legendre-submanifold}.     
%%%%%%%%%
\end{Remark}
%%%%%%%%%
%%%%%%%%%%%
\begin{Remark}
%%%%%%%%%
Consider a classical Hamiltonian system with many degrees of freedom. 
Then, the canonical equations of motion describe its dynamics in 
a $2N$-dimensional phase space with    
$N$ being assumed a large number, and $N$ is not directly related to 
the dimension of a contact manifold in general. 
To specify the most general nonequilibrium state of this Hamiltonian system, 
one needs the    
$2N$-dimensional phase space. 
Thus, when a nonequilibrium phenomenon is well-described by  
a lower dimensional contact manifold with Postulate
\ref{postulate-nonequilibrium-outside-Legendre}, 
such a nonequilibrium state is not far from equilibrium. 
%%%%%%%%%%%
\end{Remark}
%%%%%%%%%%%

The following are essential in this paper.
%%%%%%%%%%
\begin{Postulate}
%%%%%%%%%
(Time and a parameter of an integral curve) : 
Let $(\cC,\lambda)$ be a contact manifold, $h$ a contact Hamiltonian,
$X_h$ the contact Hamiltonian vector field. 
The parameter $t$ in \fr{contact-Hamiltonian-vector-components}
can physically represent time. 
%%%%%%%%%%
\end{Postulate}
%%%%%%%%%

%%%%%%%%%%%
\begin{Def}
%%%%%%%%%%%
\label{definition-relaxation-process}
(Relaxation process and attractor) : 
Let $(\cC,\lambda)$ be a contact manifold, $\cA$ a Legendre submanifold, 
$X$ a vector field on $\cC$, $t\in \mathbb{R}$ or 
$t\in T$ with some $T\subset \mathbb{R}$ parameterize 
an integral curve for $X$, and $\xi_t$ a parameterized point 
of the integral curve for $X$.
If $\xi_0\in\cC\setminus\cA$ 
and $\lim_{t\to\infty}\xi_{t}\in\cA$, then the integral curve 
is referred to as a relaxation process, and $\cA$ an attractor. 
%%%%%%%%%%%
\end{Def}
%%%%%%%%%%%
%%%%%%%%%%
\begin{Remark}
%%%%%%%%%%
A relaxation process connects a nonequilibrium state and an equilibrium state.
Geometrically an integral curve for a contact Hamiltonian vector field 
connects a point of $\cC\setminus\cA$ and that of $\cA$.
%%%%%%%%%%
\end{Remark}
%%%%%%%%%%

%%%%%%%%%%5
\begin{Postulate}
%%%%%%%%%%%%%%
(Quasi-static process and tangent vector of a Legendre submanifold) :  
An integral curve of a tangent vector field of a Legendre submanifold 
can physically represent a quasi-static process.
%%%%%%%%%%%\%%
\end{Postulate}
%%%%%%%%%%%
%%%%%%%%%%
\begin{Remark}
%%%%%%%%
In the standard thermodynamics, the speed of 
the change of equilibrium states is assumed very slow. 
However, we do not impose this.
%%%%%%%%%%
\end{Remark}
%%%%%%%%

A physical interpretation of a contact Hamiltonian
cannot be given in general.
However, some particular cases such interpretations may be given.  
For example, in Ref.\cite{Jurkowski2000}, a contact Hamiltonian is interpreted 
as a relation with the ratio of partition functions of the 
initial equilibrium system and a deformed one.  
In Ref.\cite{Bravetti-Sep-2014}, 
their contact Hamiltonian is identified with 
an entropy production potential. 
Related to this, in this paper,  
we argue that an entropy for nonequilibrium states may be  
$-\,k_{\mathrm{B}}\,\varphi$ (See \S\,\ref
{sec-statistical-manifold-legendre-submanifold}). 

%%%%%%%%%%%%%%%%
\section{Dually flat spaces, statistical manifolds, and Legendre submanifolds } 
\label{sec-statistical-manifold-legendre-submanifold}
%%%%%%%%%%%%%%
Riemannian or pseudo-Riemannian manifolds can be used to construct a  
geometric equilibrium statistical mechanics \cite{Mrugala1990,Brody2009}. 
In such equilibrium systems, 
components of a metric tensor field are expressed 
as second derivatives of a strictly convex function with respect to 
particular coordinates.  
Thus, they are related to Hessian geometry.

Since equilibrium thermodynamics is related to 
Hessian geometry and  contact geometry, and Hessian geometry is related to 
information geometry \cite{Matsuzoe2013},  
we feel that there are links among these geometries.
In this section, such links are explored.  
After some definitions and basic facts are summarized, 
it is shown how 
a contact manifold and some additional data induce 
a dually flat space used in information geometry.

%%%%%%%%%%%%%%%%%%%%%%%%%%%%%%%%%%%%%%55
\subsection{Mathematical symbols}
%%%%%%%%%%%%%%%%%%%%%%%%%%%%%%%%%%%%%%%
Mathematical symbols are fixed as follows. 

Let $(\cC,\lambda)$ be a $(2n+1)$-dimensional contact manifold,
$(x,p,z)$ the canonical coordinates such that $\lambda=\dr z-p_a\dr x^a$ 
with $x=\{x^1,\ldots,x^n\}$ and $p=\{p_1,\ldots,p_n\}$, 
$\psi\in\GamLamC{0}$ 
a function of $x$ only, $\varphi$ a function of $p$ only,  
$\Phi_{\cC\cA\psi}\,\cA_{\psi}$ the Legendre submanifold generated by $\psi$
with $\Phi_{\cC\cA\psi}:\cA_{\psi}\to\cC$ being an embedding, and 
$\Phi_{\cC\cA\varphi}\,\cA_{\varphi}$ 
the Legendre submanifold generated by $\varphi$
with $\Phi_{\cC\cA\varphi}:\cA_{\varphi}\to\cC$ being an embedding.   
These symbols and ones defined in Ref.\S\ref{sec-Mathematical-preliminaries} 
are used in \S\ref{sec-statistical-manifold-legendre-submanifold},
\S\ref{sec-attractor}, and \S\ref{sec-calculations-with-metric}.

In this section, $\psi$ and $\varphi$ are strictly convex, 
due to mathematical technicalities.   
These restrictions prevent us to describe phase transitions. 
Discussions on domains 
where $\psi$ and $\varphi$ are not strictly convex with phase transitions
are found in Ref.\cite{Bravetti-Sep-2014}. 
%%%%%%%%%%%%%%%%%%%%%%%%%%%%%%%%%%%%%%%%%%%%%%%%%%%%%%%%%%
\subsection{Legendre submanifolds in contact manifold}
%%%%%%%%%%%%%%%%%%%%%%%%%%%%%%%%%%%%%%%%%%%%%%%%%%%%%%%%%%

First a relation between Legendre submanifolds and the total Legendre 
transform of a strictly convex function is stated as follows.  
%%%%%%%%%
\begin{Lemma}
%%%%%%%%
(Legendre submanifold generated by $\psi(x)$ 
induces the one generated by $\Leg[\psi](p)$) : 
Let $\psi\in\GamLamC{0}$ be 
a strictly convex function of $x$ only, $\varphi$ the function of $p$ 
obtained by 
the total Legendre transform of $\psi$ with respect to $x$. 
Then, $\Phi_{\cC\cA\psi}\cA_{\psi}$ induces $\Phi_{\cC\cA\varphi}\cA_{\varphi}$.
%%%%%%%%%
\end{Lemma}
%%%%%%%%
%%%%%%%%%
\begin{Proof}
%%%%%%%%
At a point $\xi$ expressed as 
$(\,x,p(x),z(x)\,)\in\Phi_{\cC\cA\psi}\cA_{\psi}$,  
the equations 
$p_j=\partial\psi/\partial x^j,(j\in\{1,\ldots,n\})$ 
hold due to 
\fr{example-psi-Legendre-submanifold}.
It follows for fixed $p$ 
from Theorem\,\ref{theorem-Legendre-tranform-formula} that  
there exists the unique solution $x_*^j=x_*^j(p)$ to this equation, 
that $x_*^j=\partial\varphi/\partial p_j$, 
and that $z(x_*)=\psi(x_*)=x_*^jp_j-\varphi(p)$, for all $j\in\{1,\ldots,n\}$. 
Then, for fixed $p$, 
one can write the point $\xi\in\cC$ in terms of 
$(\,x_*(p),p,z(p)\,)$ as 
$$ 
(\,\{x_*^i(p)\},\{p_i\},z(p)\,)
=\left(\,\left\{\frac{\partial\varphi}{\partial p_i}\right\},
\{\,p_i\,\}, p_j\frac{\partial\varphi}{\partial p_j}-\varphi(p)
\,\right).
$$
So far $p$ is fixed. One then can repeat the above argument for various $p$,  
and can complete the proof.
\qed
%%%%%%%%%
\end{Proof}
%%%%%%%%

Given Legendre submanifolds 
$\Phi_{\cC\cA\psi}\cA_{\psi}$  and 
$\Phi_{\cC\cA\varphi}\,\cA_{\varphi}$ with $\varphi$ being the total Legendre 
transform of $\psi$,   
one can construct a diffeomorphism between them as follows. 
%%%%%%%%%%
\begin{Lemma}
\label{lemma-legendre-submanifolds-diffeomorphism}
%%%%%%%%%%
Let $\psi\in\GamLamC{0}$ be 
a strictly convex function of $x$ only, 
$\varphi\in\GamLamC{0}$ the function of $p$ obtained by the 
total Legendre transform of $\psi$ with respect to $x$, 
and $\Phi_{\cC\cA\psi}:\cA_{\psi}\ni x\mapsto (x,p(x),z(x))\in\cC,
\Phi_{\cC\cA\varphi}:\cA_{\varphi}\ni p\mapsto (x(p),p,z(p))\in\cC$ embeddings.    
Then, the transform  
$\cT_{\varphi\psi}:\Phi_{\cC\cA\psi}\cA_{\psi}\to\Phi_{\cC\cA\varphi}\cA_{\varphi}$  
is a diffeomorphism around 
$(\Phi_{\cC\cA\psi}\cA_{\psi})\cap(\Phi_{\cC\cA\varphi}\cA_{\varphi})\neq\emptyset$ 
( See the diagrams below ) \\
$$
\xymatrix{
\Phi_{\cC\cA\psi}\cA_{\psi}\ar[r]^{\cT_{\varphi\psi}}
& \Phi_{\cC\cA\varphi}\cA_{\varphi}
\\
\cA_{\psi}\ar[r]_{\Phi_{\varphi\psi}}\ar[u]^{\Phi_{\cC\cA\psi}}
&\cA_{\varphi} \ar[u]_{\Phi_{\cC\cA\varphi}}
}\qquad 
 \qquad 
\xymatrix{
T_{\Phi_{\cC\cA\psi}(x)}\cC\ar[r]^{\cT'_{\varphi\psi}}
&T_{\Phi_{\cC\cA\varphi}(p)}\cC \\
 T_{x}\cA_{\psi}\ar[u]^{(\Phi_{\cC\cA\psi})_*}
&T_{p}\cA_{\varphi}\ar[u]_{(\Phi_{\cC\cA\varphi})_*}
 }
$$
where 
$\Phi_{\varphi\psi}
:=\,\Phi_{\cC\cA\varphi}^{-1}\circ\cT_{\varphi\psi}\circ\Phi_{\cC\cA\psi}$ 
is also a diffeomorphism. 
%%%%%%%%%%%%%
\end{Lemma}
%%%%%%%%%%%%
%%%%%%%%%%
\begin{Proof}
%%%%%%%%%%
With \fr{example-psi-Legendre-submanifold} and 
\fr{example-varphi-Legendre-submanifold}, 
one can write 
\beqa
(\,\Phi_{\cC\cA\psi}\,)_{*}&:&
T_{x}\cA_{\psi}
\ni\quad \frac{\partial}{\partial x^j}
\mapsto 
X_j
\quad\in T_{\Phi_{\cC\cA\psi}(x)}\cC,\qquad
x\in\cA_{\psi},
\non\\
(\,\Phi_{\cC\cA\varphi}\,)_{*}&:&
T_{p}\cA_{\varphi}
\ni\quad \frac{\partial}{\partial p_i}
\mapsto 
Y^i
\quad\in T_{\Phi_{\cC\cA\varphi}(\,p\,)}\cC,\qquad
p\in\cA_{\varphi},
\non
\eeqa
around a point  
$\Phi_{\cC\cA\varphi}(\,p\,)=\Phi_{\cC\cA\psi}(\,x\,)=(x,p,z)$, 
where
\beqa
X_j
&:=&(\,\Phi_{\cC\cA\psi}\,)_{*}\left(\frac{\partial}{\partial x^j}\right)
=\frac{\partial}{\partial x^j}
+\frac{\partial p_b}{\partial x^j}\frac{\partial}{\partial p_b}
+\frac{\partial z}{\partial x^j}\frac{\partial}{\partial z}
=\frac{\partial}{\partial x^j}
+\frac{\partial^2 \psi}{\partial x^j\partial x^b}\frac{\partial}{\partial p_b}
+\frac{\partial \psi}{\partial x^j}\frac{\partial}{\partial z},
\non\\
Y^i
&:=&(\,\Phi_{\cC\cA\varphi}\,)_{*}\left(\frac{\partial}{\partial p_i}\right)
=\frac{\partial x^b}{\partial p_i}\frac{\partial}{\partial x^b}
+\frac{\partial}{\partial p_i}
+\frac{\partial z}{\partial p_i}\frac{\partial}{\partial z}
=\frac{\partial^2 \varphi}{\partial p_i\partial p_b}
\frac{\partial}{\partial x^b}
+\frac{\partial}{\partial p_i}
+p_b\frac{\partial^2 \varphi}{\partial p_i\partial p_b}
\frac{\partial}{\partial z}.
\non
\eeqa
Define 
$$
S^{ab}
:=\frac{\partial^2\varphi}{\partial p_a\partial p_b}
=S^{ba},
$$
where $\det(S)\neq 0$ due to $\varphi$ being a strictly convex function.
With  
$$
\frac{\partial^2 \varphi}{\partial p_a\partial p_b}
\frac{\partial^2\psi}{\partial x^b\partial x^c}
=\frac{\partial x^a}{\partial p_b}
\frac{\partial p_b}{\partial x^c}
=\delta_c^a\quad\mbox{and}\quad
\frac{\partial\psi}{\partial x^b}
=p_b,
$$
one has 
\beqa
S^{ab}X_b
&=&\frac{\partial^2\varphi}{\partial p_a\partial p_b}
\left(\,
\frac{\partial}{\partial x^b}
+\frac{\partial^2 \psi}{\partial x^b\partial x^c}\frac{\partial}{\partial p_c}
+\frac{\partial \psi}{\partial x^b}\frac{\partial}{\partial z}
\,\right)
\non\\
&=&\frac{\partial^2 \varphi}{\partial p_a\partial p_b}
\frac{\partial}{\partial x^b}
+\frac{\partial^2 \varphi}{\partial p_a\partial p_b}
\frac{\partial^2\psi}{\partial x^b\partial x^c}
\frac{\partial}{\partial p_c}
+\frac{\partial\psi}{\partial x^b}
\frac{\partial^2 \varphi}{\partial p_a\partial p_b}
\frac{\partial}{\partial z}
\non\\
&=&\frac{\partial^2 \varphi}{\partial p_a\partial p_b}
\frac{\partial}{\partial x^b}
+\frac{\partial}{\partial p_a}
+p_b\frac{\partial^2 \varphi}{\partial p_a\partial p_b}
\frac{\partial}{\partial z}
=Y^a,
\non
\eeqa
from which $Y^a=S^{ab}X_b$.

Since $\det (S)\neq 0$, the linear map 
$$
\cT'_{\varphi\psi} : T_{\Phi_{\cC\cA\psi}(x)}\cC\ni \{X_j\}\,
\mapsto \,
\{Y^i\}\in T_{\Phi_{\cC\cA\varphi}(p)}\cC,\qquad 
Y^a=S^{ab}X_b,
$$  
is isomorphic.
Applying the inverse function theorem  with this map 
$\cT'_{\varphi\psi} : T_{\Phi_{\cC\cA\psi}(x)}\cC \to T_{\Phi_{\cC\cA\varphi}(p)}\cC$, 
one concludes that the map 
$\cT_{\varphi\psi}: \Phi_{\cC\cA\psi}\cA_{\psi}\to \Phi_{\cC\cA\varphi}\cA_{\varphi}$  
is diffeomorphic around 
$\Phi_{\cC\cA\varphi}(\,p\,)=\Phi_{\cC\cA\psi}(x)=(x,p,z)\in\cC$. 
In addition, since 
$\Phi_{\cC\cA\psi}$ and $\Phi_{\cC\cA\varphi}$ are one-to-one due to   
$\cA_{\psi}$ and $\cA_{\varphi}$ being submanifolds, 
$\Phi_{\varphi\psi}
:=\,\Phi_{\cC\cA\varphi}^{-1}\circ\cT_{\varphi\psi}\circ\Phi_{\cC\cA\psi}$ 
is also a diffeomorphism. 
\qed
%%%%%%%%%%
\end{Proof}
%%%%%%%%%%
%%%%%%
\begin{Remark}
%%%%%%%%
The idea of the above proof is based on Ref.\cite{Bravetti-Sep-2014}.
%%%%%%
\end{Remark}
%%%%%%%%

%%%%%
\begin{Thm}
%%%%
\label{theorem-contact-manifold-convex-function-induce-Riemannian}
(Contact manifold and a strictly convex function induce  
a Riemannian manifold on a Legendre submanifold) :
Let $\psi\in\GamLamC{0}$ be a strictly convex function of $x$ only, and 
$\varphi\in\GamLamC{0}$ a strictly convex function of $p$ only. 
Define 
$$
g^{\,\cA\psi}
:=g_{ab}^{\,\cA\psi}\,\dr \theta^a\otimes \dr\theta^b\in\GamLamA{2}_{\psi},
\qquad 
g_{ab}^{\,\cA\psi}
:=\Phi_{\cC\cA\psi}^{\ \ \ \ *}\frac{\partial^2\psi}{\partial x^a\partial x^b},
\quad
\theta^a
:=\Phi_{\cC\cA\psi}^{\ \ \ \ *}x^a,
$$ 
and 
$$
g^{\,\cA\varphi}
:=g^{\,\cA\varphi\ ab}\,\dr \eta_a\otimes \dr\eta_b\in\GamLamA{2}_{\varphi},
\qquad 
g^{\,\cA\varphi\ ab}
:=\Phi_{\cC\cA\varphi}^{\ \ \ \ *}\frac{\partial^2\varphi}{\partial p_a\partial p_b},
\quad
\eta_a
:=\Phi_{\cC\cA\varphi}^{\ \ \ \ *}p_a,
$$ 
then $(\cA_{\psi},g^{\,\cA\psi})$ and $(\cA_{\varphi},g^{\,\cA\varphi})$ are 
$n$-dimensional Riemannian manifolds.
%%%%%
\end{Thm}
%%%%
%%%%
\begin{Proof}
%%%%
It follows from  $\psi$ and $\varphi$ 
being strictly convex functions that 
$\det(\partial^2\psi/\partial x^a\partial x^b)\neq 0$, and\\ 
$\det(\partial^2\varphi/\partial p_a\partial p_b)\neq 0$.
\qed
%%%%5
\end{Proof}
%%%%
If $\varphi$ in 
Theorem\,\ref{theorem-contact-manifold-convex-function-induce-Riemannian} 
is obtained by the total Legendre transform of 
$\psi$ with respect to $x$, then 
it is shown in the following that 
the inverse matrix of $\{g_{ab}^{\,\cA\psi}\}$ 
is concisely written in terms of the derivatives of $\varphi$.

%%%%%%%%%%
\begin{Lemma}
%%%%%%%%%%
\label{lemma-legendre-submanifolds-diffeomorphism-inverse}
Let $\psi\in\GamLamC{0}$ be a strictly convex function 
of $x$ only, $\varphi\in\GamLamC{0}$ the function of $p$ obtained by 
 the total Legendre transform of $\psi$ with respect to $x$, 
and $\Phi_{\cC\cA\psi}:\cA_{\psi}\to\cC$ and 
$\Phi_{\cC\cA\varphi}:\cA_{\varphi}\to\cC$ embeddings ( See the diagrams below ).   
Define $\{g_{ab}^{\,\cC}\}$, and $\{g^{\,\cC\ ab}\}$  to be  
\beq
g_{ab}^{\,\cC}
:=\frac{\partial^2\psi}{\partial x^a\partial x^b},\ \in\GamLamC{0}\qquad
g^{\,\cC\ ab}
:=\frac{\partial^2\varphi}{\partial p_a\partial p_b},\ \in\GamLamC{0}.
\label{metric-on-Legendre-submanifolds-from-psi-varphi}
\eeq

Then, around a point where 
$(\Phi_{\cC\cA\psi}\cA_{\psi})\cap(\Phi_{\cC\cA\varphi}\cA_{\varphi})\neq \emptyset$, 
one has 
$$
(\,\Phi_{\cC\cA\psi}^{\ \ \ \ *}g_{al}^{\,\cC}\,)
\left[\,\Phi_{\varphi\psi}^{\ \ \ *}
(\,\Phi_{\cC\cA\varphi}^{\ \ \ \ *}g^{\,\cC\ lb}\,)\,\right]
=\delta_a^b,
$$
where $\Phi_{\varphi\psi}: \cA_{\psi}\to\cA_{\varphi}$. 
$$
\xymatrix{
\Phi_{\cC\cA\psi}\cA_{\psi}
& \Phi_{\cC\cA\varphi}\cA_{\varphi}
\\
\cA_{\psi}
\ar[r]_{\Phi_{\varphi\psi}}
\ar[u]^{\Phi_{\cC\cA\psi}}
&\cA_{\varphi} 
\ar[u]_{\Phi_{\cC\cA\varphi}}
}\qquad \qquad
\xymatrix{
\Gt{0}{q}{\Phi_{\cC\cA\psi}\cA_{\psi}} 
\ar[d]_{\Phi_{\cC\cA\psi}^{\ \ \ \ *}}
&\Gt{0}{q}{\Phi_{\cC\cA\varphi}\cA_{\varphi}}
\ar[d]^{\Phi_{\cC\cA\varphi}^{\ \ \ \ *}} \\
 \Gt{0}{q}{\cA_{\psi}}
&\Gt{0}{q}{\cA_{\varphi}}
\ar[l]^{\Phi_{\varphi\psi}^{\ \ \ *}}
 }
$$
%%%%%%%%%%%%
\end{Lemma}
%%%%%%%%%%
%%%%%%%%%
\begin{Proof}
%%%%%%%%
It follows from 
\fr{example-psi-Legendre-submanifold},  
and \fr{example-varphi-Legendre-submanifold}  
that
$$
(\,\Phi_{\cC\cA\psi}^{\ \ \ \ *}g_{al}^{\,\cC}\,)
\left[\,\Phi_{\varphi\psi}^{\ \ \ *}
  (\,\Phi_{\cC\cA\varphi}^{\ \ \ \ *}g^{\,\cC\, lb}\,)\,\right]
=\left(\,
\Phi_{\cC\cA\psi}^{\ \ \ \ *}\frac{\partial^2\psi}{\partial x^a\partial x^l}
\,\right)\,\left[\,\Phi_{\varphi\psi}^{\ \ *}
\left(\,
\Phi_{\cC\cA\varphi}^{\ \ \ \ *}\frac{\partial^2\varphi}{\partial p_l\partial p_b}
\,\right)\,\right]
=\frac{\partial p_a}{\partial x^l}\,
\frac{\partial x^l}{\partial p_b}
=\delta_a^b.
$$
\qed
%%%%%%%%%
\end{Proof}
%%%%%%%%
%%%%%%%%%
\begin{Remark}
%%%%%%%%
This lemma states that the pull-back of 
$\{g_{ab}^{\,\cC}\}\in\GamLamC{0}$ and that of $\{g^{\,\cC\ ab}\}\in\GamLamC{0}$ 
can be used as components of a metric tensor field  
on a Legendre submanifold of $\cC$.
%%%%%%%%%
\end{Remark}
%%%%%%%%

%%%%%%%%%%%%
\subsection{Relations among dually flat spaces, statistical manifolds, and Legendre submanifolds}
%%%%%%%%%%%%\
To say a relation between a Legendre submanifold of a contact manifold and 
a dually flat space or statistical manifold, 
one needs the following definitions. 

%%%%%%%%%
\begin{Def}
%%%%%%%%%
(Affine-coordinate and flat connection, \cite{AN}) : 
Let $\cM$ be an $n$-dimensional manifold, $\theta:=\{\theta^1,\ldots,\theta^n\}$
coordinates, $\nabla$ a connection, $\{\Gamma_{ab}^{\ \ c}\}$ 
connection coefficients such that 
$\nabla_{\partial_a}\partial_b=\Gamma_{ab}^{\ \ c}\partial_c,$ 
$(\partial_a:=\partial/\partial\theta^a)$. If 
$\{\Gamma_{ab}^{\ \ c }\}\equiv 0$ hold for all $\xi\in\cM$, then
$\theta$ is referred to as a $\nabla$-affine coordinate system, 
or simply affine coordinates. 
If it is the case, then $\nabla$ 
is referred to as a flat connection. 
%%%%%%%%%
\end{Def}
%%%%%%%%%
%%%%%%%%%
\begin{Def}
%%%%%%%%%
(Hessian manifold, \cite{Matsuzoe2013}) : 
Let $(\cH,g)$ be an $n$-dimensional Riemannian or 
pseudo-Riemannian manifold, $\nabla$ a connection, and 
$\Psi$ a strictly convex function on $\cH$. 
If (i) $g=\nabla\dr\Psi$ holds, and (ii) $\nabla$-affine coordinates exist, 
then $(\cH,\nabla,\nabla\dr\Psi)$ is referred to as a Hessian manifold or a 
Hesse manifold.
%%%%%%%%%
\end{Def}
%%%%%%%%%
%%%%%%%%%%%%%%%%
\begin{Remark}
%%%%%%%%%%%%%%%%%%%
Let $\cM$ be an $n$-dimensional manifold, $\nabla'$ a connection, $\theta'$ 
a coordinate system which is not necessary to be $\nabla'$-affine, 
and $\Psi'$ a strictly convex function.
Then, the components of 
$h':=\nabla'\dr\Psi^{\,\prime}
=h_{ab}^{\prime}\,\dr\theta^{\,\prime a}\otimes\dr\theta^{\,\prime b}$ are written as 
$$
h_{ab}^{\,\prime}
=\frac{\partial^2\Psi^{\,\prime}}{\partial\theta^{\,\prime a}\partial\theta^{\,\prime b}}
-\Gamma_{ab}^{\,\prime \ c}\frac{\partial\Psi^{\,\prime}}{\partial\theta^{\,\prime c}}.
$$
For an $n$-dimensional 
Hessian manifold $(\cH,\nabla,\nabla\dr\Psi)$ with $g=\nabla\dr\Psi$, 
the components of $g=g_{ab}\,\dr\theta^{\,a}\otimes\dr\theta^{\,b}$  
with $\nabla$-affine coordinates $\theta$ are
$$
g_{ab}
=\frac{\partial^2\Psi}{\partial\theta^a\partial\theta^b}.
$$
Since $\Psi$ is a strictly convex function, there exists the inverse matrix of 
$(\,g_{ab}\,)$.
%%%%%%%%%%%%%
\end{Remark}
%%%%%%%%%%
%%%%%%%%%
\begin{Def}
%%%%%%%%%
(Dual connection, \cite{AN}) : 
Let $(\cM,g)$ be an $n$-dimensional Riemannian or pseudo-Riemannian manifold,
$\nabla$ and $\nabla^*$ connections. If 
$$
X\,\left[\,g(\,Y,Z\,)\,\right]
=g(\,\nabla_XY,Z\,)+g(\,Y,\nabla_X^{*}Z\,),\qquad \forall\, X,Y,Z\in\GTM
$$
then $\nabla$ and $\nabla^*$ are referred to as dual connections, also 
$\nabla^*$ is referred to as a dual connection of $\nabla$ with respect to $g$. 
%%%%%%%%%
\end{Def}
%%%%%%%%%
%%%%%%%%%
\begin{Def}
%%%%%%%%%
(Dually flat space, \cite{AN}) : 
Let $(\cM,g)$ be an $n$-dimensional Riemannian or pseudo-Riemannian manifold, 
$\nabla$ and $\nabla^*$ dual connections, then $(\cM,g,\nabla,\nabla^*)$ is 
referred to as a dually flat space. 
%%%%%%%%%
\end{Def}
%%%%%%%%%
%%%%%%%%%
\begin{Def}
%%%%%%%%%
(Statistical manifold, \cite{Matsuzoe2013}) : 
Let $(\cM,g)$ be an $n$-dimensional Riemannian or pseudo-Riemannian manifold,
$\nabla$ a torsion-free connection. If $\nabla g$ is symmetric, 
then $(\cM,\nabla,g)$ is referred to as an $n$-dimensional statistical 
manifold. 
%%%%%%%%%
\end{Def}
%%%%%%%%%

A metric tensor field $g$ on a statistical manifold is related to 
the following. 
%%%%%%%%%
\begin{Def}
%%%%%%%%%
(Fisher information matrix, \cite{AN}) : 
Let $\zeta$ be a set of random variables, 
$\theta:=\{\theta^1,\ldots,\theta^n\}\in\Theta$ some parameters, 
$\mbbP_{\theta}$ a distribution function parameterized by $\theta$, 
and $f$ a function of $\zeta$ and $\theta$. 
Then, with 
\beq
g_{ab}^{\F}(\theta)
:=\mbbE_{\theta}\left[\frac{\partial\,\ln\mbbP_{\theta}}{\partial\theta^a}
\frac{\partial\,\ln\mbbP_{\theta}}{\partial\theta^b}\right],
\quad a,b\in\{1,\ldots,n\},
\qquad
\mbbE_{\theta}[f]
:=\int\dr\zeta\,\mbbP_{\theta}\,f(\,\zeta,\theta\,),
\label{definition-Fisher-matrix}
\eeq
%%%%%%%%%%
the matrix $(\,g_{ab}^{\F}\,)$ 
is referred to as the Fisher information matrix.
%%%%%%%%%
\end{Def}
%%%%%%%%%

A connection on a statistical manifold is related to the 
following one-parameter family of connections. 
%%%%%%%%%
\begin{Def}
%%%%%%%%%
(\,$\alpha$-connection, \cite{AN}) : 
Let $\zeta$ be a set of random variables,  
$\theta:=\{\theta^1,\ldots,\theta^n\}\in\Theta$ some parameters,
$\mbbP_{\theta}$ a distribution function parameterized by $\theta$, 
$\alpha$ a real number, and $f$ a 
function of $\zeta$ and $\theta$. 
Then, with 
$$ 
\Gamma_{abc}^{(\alpha)}
:=\mbbE_{\theta}\left[\left(\frac{\partial^2\,\ln\mbbP_{\theta}}{\partial\theta^a\partial\theta^b}+\frac{1-\alpha}{2}\frac{\partial^2\,\ln\mbbP_{\theta}}{\partial\theta^a\partial\theta^b}\right)
\left(\frac{\partial\,\ln\mbbP_{\theta}}{\partial\theta^c}\right)\right],
\quad a,b,c\in\{1,\ldots,n\},
\quad
\mbbE_{\theta}[f]
:=\int\dr\zeta\,\mbbP_{\theta}\,f(\,\zeta,\theta\,),
$$
and 
$$
\nabla_{\partial_a}^{(\alpha)}\partial_b
=\Gamma_{ab}^{(\alpha)\ c}\partial_c, 
\qquad 
\Gamma_{ab}^{(\alpha)\ c}
:=g^{\F\,cj}\Gamma_{abj}^{(\alpha)},\qquad
\partial_a
:=\frac{\partial}{\partial\theta^a},
$$
the connection $\nabla^{(\alpha)}$ 
is referred to as the $\alpha$-connection.
%%%%%%%%%
\end{Def}
%%%%%%%%%

The following example shows how a distribution function is connected to    
a Hessian manifold and a dually flat space.
%%%%%%%
\begin{Example}
%%%%%%% 
(Exponential family, \cite{AN}) :  
Consider the set of the exponential family
$\cS=\{\,\mbbP_{\theta}(\zeta)\,\}$ 
where $\zeta$ is a set of random variables, and  
$\mbbP_{\theta}(\zeta)$ a probability distribution function 
parameterized by $\theta=\{\theta^1,\ldots,\theta^n\}\in\Theta$ 
as 
$$
\mbbP_{\theta}(\zeta)
=\exp\left(\,C(\zeta)+\theta^a F_a(\zeta)-\psi^{\,\cS}(\theta)\,\right).
$$
Here, $C$ and $\{F_a\}$ are functions of $\zeta$, 
$\psi^{\,\cS}$ is referred to as the cumulant generating function 
that is to normalize $\mbbP_{\theta}$. 
The explicit form of $\psi^{\,\cS}(\theta)$ is obtained as 
$$
\psi^{\,\cS}(\theta)
=\ln\left[
\int\dr \zeta\,\exp\left(\,C(\zeta)+\theta^aF_a(\zeta)\,\right)
\right],
$$
which is convex. 
It is straightforward to show that the components of the 
Fisher matrix defined in  \fr{definition-Fisher-matrix} are  
$$
g_{ab}^{\F}
=\frac{\partial^2\psi^{\cS}}{\partial\theta^a\partial\theta^b},
$$
and $g^{\F}:=g_{ab}^{\F}\dr\theta^a\otimes\dr\theta^b$ becomes a metric tensor field for the domain where $\det(g_{ab}^{\F})>0$ is satisfied. 
In the following $\det(g_{ab}^{\F})>0$ is assumed.  
Then, one has \cite{Matsuzoe2013} 
%%%%%%%%%%%%%%%%
\begin{itemize}
%%%%%%%%%%%%%%%
%%%%%%
\item
%%%%%
$(\cS,\nabla^{(\alpha)},g^{\F})$ is a statistical manifold. 
%%%%%%%
\item
%%%%%
$(\cS,\nabla^{(1)},g^{\F})$ is a Hessian manifold where 
$\nabla^{(1)}$-affine coordinates are $\{\theta^a\}$, 
%%%%%%
\item
%%%%%%
$(\cS,g^{\F},\nabla^{(1)},\nabla^{(-1)})$ is a dually flat space where 
$\nabla^{(-1)}$-affine coordinates are $\eta:=\{\eta_1,\ldots,\eta_n\}$ 
with 
$\eta_a:=\partial\psi^{\,\cS}/\partial\theta^a$.
%%%%%%%%%%%%%
\end{itemize}
%%%%%%%%%%%%
It is worth noting the explicit form of $\varphi^{\,\cS}:=\Leg[\psi^{\,\cS}]$. 
It follows that 
$$
\varphi^{\,\cS}(\eta)
=\mbbE_{\theta}\left[\,\ln \mbbP_{\theta}\,\right](\eta)
-\mbbE_{\theta}\left[\,C\,\right](\eta),\qquad\mbox{where}\quad 
\mbbE_{\theta}\left[\,f\,\right]
:=\int\dr\zeta\, \mbbP_{\theta}(\zeta)\,f(\zeta,\theta),
$$ 
with $f$ being an arbitrary function. 
%%%%%%%
\end{Example}
%%%%%%%

The following example shows how the {exponential family} is used in 
geometrization of equilibrium statistical mechanics.
%%%%%%%
\begin{Example}
%%%%%%% 
\label{example-grand-canonical-distribution}
(Grand canonical distribution) : 
Consider the grand canonical distribution where 
a probability distribution function $\mbbP_{\theta}\in\cS$   
parameterized by $\theta^1=-1/(k_{\B}T_{\abs})=-\beta_{\abs},
\theta^2=\mu_1/(k_{\B}T_{\abs})=\mu_1\beta_{\abs}$, 
and is of the form 
$$
\mbbP_{\theta}(\zeta)
=\exp\left(\,\theta^aF_a(\zeta)-\psi^{\,\cS}(\theta)\,\right)
=\exp\left(\,-\,\frac{(\,H_N(\zeta)-\mu_1N(\zeta)\,)}{k_{\B}T_{\abs}} 
-\ln Z_{\G}(T_{\abs},\mu_1)\,\right),
$$
where $F_1(\zeta)=H_N(\zeta)$ is a Hamiltonian at a micro-state $\zeta$, 
$F_2(\zeta)=N(\zeta)$ the number of the particles at $\zeta$, 
and $\psi^{\,\cS}(\theta)=\ln Z_{\G}(T_{\abs},\mu_1)$ with $Z_{\G}$ being 
the so-called grand partition function to normalize 
$\mbbP_{\theta}$. 
Observe that this distribution function belongs to the exponential family.
It is worth noting the explicit 
forms of $\psi^{\cS}$ and $\varphi^{\cS}$. It follows that  
$$
\psi^{\cS}(\theta)
=\ln\left[\,\int\dr\zeta\,\exp
\left(\,-\,\frac{H_N(\zeta)-\mu_1N(\zeta)}{k_{\B}T_{\abs}}\,\right)\right]
=\ln Z_{\G}(T_{\abs},\mu_1)
=-\,\frac{1}{k_{\B}T_{\abs}}\Omega_{\G}\left(T_{\abs},\mu_1\right),
$$
where 
$\Omega_{\G}(T_{\abs},\mu_1)=-\,k_{\B}T_{\abs}\ln Z_{\G}(T_{\abs},\mu_1)$ 
is the grand canonical potential that reduces to $\cF$ called the 
Helmholtz free energy
when $\mu_1=0$. With $\psi^{\cS}$, one obtains 
$$
\eta_1
=\frac{\partial\psi^{\cS}}{\partial \theta^1}
=\mbbE_{\theta}[H_N]
=:\ave{H_N},\quad
\eta_2
=\frac{\partial\psi^{\cS}}{\partial \theta^2}
=\mbbE_{\theta}[N]
=:\ave{N},
\quad\mbox{where}\quad
\mbbE_{\theta}[f]
:=\int\dr\zeta\,\mbbP_{\theta}(\zeta)f(\zeta,\theta),
$$
with $f$ being arbitrary function. 
The total Legendre transform 
of $\psi^{\cS}$ with respect to $\theta$ is expressed as 
$$
\varphi^{\cS}(\eta)
=\theta_*^1\eta_1+\theta_*^2\eta_2-\psi(\theta_*)
=-\,\frac{\ave{H_N}}{k_{\B}T_{\abs\,*}}+\frac{\mu_{1\,*}\ave{N}}{k_{\B}T_{\abs\,*}}
+\frac{\Omega_G(T_{\abs\,*},\mu_{1\,*})}{k_{\B}T_{\abs\,*}}
=-\,\frac{S(\ave{H_N},\ave{N})}{k_{\B}},
$$
where $T_{\abs\,*}=-1/(k_{\B}\theta_*^1)$ and $\mu_{1\,*}=k_{\B}T_{\abs\,*}\theta_*^2$ 
with 
$\theta_*^1=\theta_*^1(\eta)$ and $\theta_*^2=\theta_*^2(\eta)$
being the solutions to 
$$
\eta_1
=\left.\frac{\partial\psi^{\cS}}{\partial\theta^1}\right|_{\theta=\theta_*},
\qquad
\eta_2
=\left.\frac{\partial\psi^{\cS}}{\partial\theta^2}\right|_{\theta=\theta_*},
$$
respectively, 
and the relation $\Omega_G=\ave{H_N}-\mu_1\ave{N}-T_{\abs}S$, known in 
equilibrium thermodynamics, has been used.  
%%%%%%%
\end{Example}
%%%%%%% 

In addition to the exponential family, 
a deformed exponential family is also linked to 
a Hessian manifold and a dually flat space \cite{Matsuzoe2013}.  

Hessian manifolds are related to dually flat spaces as follows. 
%%%%%%%%
\begin{Proposition}
%%%%%%%%%
\label{hessian-manifold-induce-dually-flat-space}
(A Hessian manifold induces a dually flat space) : 
Let $(\cH,\nabla,\nabla\dr\Psi)$ be a 
Hessian manifold with $g=\nabla\dr\Psi$. If $\cH$ is simply connected,  
then $(\cH,\nabla,\nabla\dr\Psi)$ induces %the dually flat space 
$(\cH,\nabla^*,\nabla^*\dr\,\Psi^{\cL})$, where $\Psi^{\cL}$ 
is the total Legendre transform 
of $\Psi$ with respect to $\nabla$-affine coordinates, and 
$\nabla^*$  a dual connection of 
$\nabla$ with respect to $g$. 
%%%%%%%%
\end{Proposition}
%%%%%%%%%
\begin{Proof}
%%%%%%%%%%%%%
A proof is based on the proof of Theorem\,3.6 in Ref.\cite{AN}.
\qed
%%%%%%%%%%
\end{Proof}
%%%%%%%%%%%%

The following is a key to 
connect a dually flat space %statistical manifold 
and a contact manifold. 
%%%%%%%%%%
\begin{Proposition}
\label{contact-function-induce-hessian-manifold}
%%%%%%%%%
(A contact manifold and a strictly convex function 
induce a Hessian manifold ) : 
Let $(\cC,\lambda)$ be a $(2n+1)$-dimensional contact manifold, 
$(x,p,z)$ canonical coordinates such that $\lambda=\dr z-p_a\dr x^a$ 
with $x=\{x^1,\ldots,x^n\}$ and $p=\{p_1,\ldots,p_n\}$, and 
$\psi$ a strictly convex function of $x$ only.
Then, $(\,(\cC,\lambda),\psi\,)$ induces the $n$-dimensional Hessian manifold 
$(\cH,\nabla,\nabla\dr\psi)$. 
%%%%%%%%%%
\end{Proposition}
%%%%%%%%%
%%%%%%%%
\begin{Proof}
%%%%%%%
Let $\Phi_{\cC\cA\psi}\cA_{\psi}$ be the Legendre submanifold generated by $\psi$  
with $\Phi_{\cC\cA\psi}:\cA_{\psi}\to\cC$ being an embedding, 
and $\theta^a:=\left.x^a\right|_{\Phi\cA_{\psi}}$, $(a\in\{1,\ldots,n\})$.  
Then, it follows from 
$\det\,(\,\partial^2\psi/\partial x^a\partial x^b\,)>0,(\forall\xi\in\cC)$  
that 
$g_{ab}:=\Phi^*(\,\partial^2\psi/\partial x^a\partial x^b\,)$
can be used as components of a metric tensor field on $\cA_{\psi}$. Thus, 
identifying $\cH=\Phi\cA_{\psi}$, $\nabla$-affine coordinates 
to be $\{\theta^a\}$, and 
$\nabla\dr\psi=\Phi^*(\,\partial^2\psi/\partial x^a\partial x^b\,)\,\dr 
\theta^a\otimes\dr \theta^b$, one has the Hessian manifold.
\qed
%%%%a%%%%
\end{Proof}
%%%%%%%

The following theorem is the main claim in this section, and it stipulates a 
relation between 
a dually flat space and a contact manifold. 
%%%%%%%%%%
\begin{Thm}
\label{contact-induce-dually-flat-space}
%%%%%%%%%
(A contact manifold and a strictly convex function 
induce a dually flat space ) : 
Let $(\cC,\lambda)$ be a $(2n+1)$-dimensional contact manifold, 
$(x,p,z)$ canonical coordinates such that $\lambda=\dr z-p_a\dr x^a$ 
with $x=\{x^1,\ldots,x^n\}$ and $p=\{p_1,\ldots,p_n\}$, and 
$\psi$ a strictly convex function of $x$ only. If the Legendre submanifold 
generated by $\psi$ is simply connected, 
then $(\,(\cC,\lambda),\psi\,)$ induces the $n$-dimensional dually flat space 
$(\cH,g,\nabla,\nabla^*)$. 
%%%%%%%%%%
\end{Thm}
%%%%%%%%%
%%%%%%%%
\begin{Proof}
%%%%%%%
First, it follows from 
Proposition\,\ref{contact-function-induce-hessian-manifold} 
that $(\,(\cC,\lambda),\psi\,)$ induces 
$(\cH,\nabla,\nabla\dr\psi)$. Second, it follows from 
Proposition\,\ref{hessian-manifold-induce-dually-flat-space} 
that $(\cH,\nabla,\nabla\dr\psi)$ induces 
$(\cH,\nabla^*,\nabla^*\dr\,\varphi)$, where $\varphi$ is the total 
Legendre transform of $\psi$ with respect to $\nabla$-affine coordinates. 
Thus,  
one has the dual connection $\nabla^*$ on $\cH$.  
Combining these, one has 
that $(\,(\cC,\lambda),\psi\,)$ induces 
$(\cH,g,\nabla,\nabla^*)$.
\qed
%%%%%%%%
\end{Proof}
%%%%%%%
%%%%%%%
\begin{Remark}
%%%%%%%%%%%%%%%
In Theorem\,\ref{contact-induce-dually-flat-space},  
if $\nabla$ is torsion-free and $\nabla g$ is symmetric, then 
$(\,(\cC,\lambda),\psi\,)$ induces the statistical manifold 
$(\cH,\nabla,g)$.
%%%%%%%%%%%%
\end{Remark}
%%%%%%5\
%%%%%%%
\begin{Remark}
%%%%%%%%%%%%%%%
Similar to Theorem\,\ref{contact-induce-dually-flat-space},  
one can show that 
$(\,(\cC,\lambda),\varphi\,)$ induces the $n$-dimensional dually flat space 
$(\cH,g,\nabla,\nabla^*)$.
%%%%%%%%%%%%
\end{Remark}
%%%%%%5\

It is interesting to seek a prescription that gives a $(2n+1)$-dimensional 
contact manifold $(\cC,\lambda)$ from a given $n$-dimensional 
Hessian manifold 
 $(\cA,\nabla,\nabla\dr\Psi)$.   
To our knowledge such a prescription has not been known. 
Consider Example 
\ref{example-grand-canonical-distribution}. 
If such a prescription is found, then 
$-\, k_{\B}\,\varphi\in\GamLamC{0}$  may be 
an entropy 
for nonequilibrium states, where $\varphi$ is obtained 
by the total Legendre transform of   
a function $\psi\in\GamLamC{0}$, 
and $\psi$ is obtained from $\Psi\in\GamLamA{0}$ such that 
$\psi|_{\cA}=\Psi$.

%%%%%%%%%%%
\section{ Legendre submanifolds as attractors in contact manifold}
\label{sec-attractor}
%%%%%%%%%%55
Of particular interest for this paper is contact Hamiltonian 
vector fields whose integral curves are 
relaxation processes in a contact manifold.

%%%%%%%%%%%%%%%%
\subsection{General theory}
%%%%%%%%%%%%%%%\
The main claim in this section is the following.
Roughly speaking the following theorem states that there is a class of 
contact Hamiltonians such that 
the Legendre submanifold generated by a given function  
becomes an attractor of the contact Hamiltonian vector field. 

%%%%%%%%%%%%%%
\begin{Thm}
\label{theorem:main-atrractor-psi-1}
%%%%%%%%%%%%
(Relaxation process in terms of contact Hamiltonian vector field $1$) : 
Let $\psi\in\GamLamC{0}$ be a function of $x$ only.
Define $h_{\psi}, \Delta_{\psi}\in\GamLamC{0}$, $\cD_{\psi}\subset \cC$ and 
$\wh{h}:\mathbb{R}\to\mathbb{R}$ such that   
$$
h_{\psi}(x,z)
:=\wh{h}(\Delta_{\psi}),\quad 
\Delta_{\psi}(x,z)
:=\psi(x)-z,
$$
$$
\cD_{\psi}
:=\left\{\,(x,p,z)\in\cC\,
\bigg|\,\wh{h}(\Delta_{\psi})\geq 0\quad\mbox{and}\quad\frac{\dr \wh{h}}{\dr\Delta_{\psi}}>0
\,\right\}\subset \cC,
$$
and 
$$
\wh{h}(0)
=0,\qquad\
\wh{h}(\Delta_{\psi})>0,\ \mbox{for}\ \Delta_{\psi}\neq 0.
$$
Then, integral curves of the contact vector field associated to $h_{\psi}$ 
on $\cD_{\psi}$ 
connect points of $\cD_{\psi}$ and those of 
$\Phi_{\cC\cA\psi}\cA_{\psi}$.  
Thus, the integral curves can be 
relaxation processes (See Definition \ref{definition-relaxation-process}). 
%%%%%%%%%
\end{Thm}
%%%%%%%%
%%%%%%%%%%%%%%%
\begin{Proof}
%%%%%%%%%%%%%%
The contact Hamiltonian vector field $X_h^{\psi}$ 
is the flow that can be expressed by 
\fr{contact-Hamiltonian-vector-components},  
\beq
X_h^{\psi}
=\dot{x}^j\frac{\partial}{\partial x^j}
+\dot{p}_j\frac{\partial}{\partial p_j}
+\dot{z}\frac{\partial}{\partial z},\qquad
\frac{\dr x^j}{\dr t}
=0,
\qquad 
\frac{\dr p_j}{\dr t}
=\left(\,\frac{\partial\psi}{\partial x^j}-p_j\,\right)
\,\frac{\dr \wh{h}}{\dr \Delta_{\psi}}, 
\qquad 
\frac{\dr z}{\dr t}
=\wh{h}.
\label{contact-Hamiltonian-Lyapunov-psi-vector}
\eeq
The set of fixed points of this dynamical system 
on $\cD_{\psi}\subset \cC$ is 
found to be 
$$
\cF_{\psi}
=\left\{\ (\,\ol{x},\ol{p},\,\ol{z}\,)\in\cC \ \bigg|\ 
\frac{\partial\psi}{\partial \ol{x}^j}-\ol{p}_j
=0,\ \mbox{and}\ 
\Delta_{\psi}(\,\ol{x},\ol{z}\,)
=\psi(\,\ol{x}\,)-\ol{z}
=0
\right\}.
$$
On the other hand the Legendre submanifold $\cA_{\psi}$ is given by 
\fr{example-psi-Legendre-submanifold}. 
Introduce the abbreviation for $\Phi_{\cC\cA\psi}\,\cA_{\psi}$ with 
$\Phi_{\cC\cA\psi}:\cA_{\psi}\to\cC$ as
$$
\cA_{\psi}^{\cC}
:=\Phi_{\cC\cA\psi}\cA_{\psi}
:=\left\{\ (\ x^{(\cA)},p^{(\cA)},z^{(\cA)}\ )\in\cC\ \bigg|\ 
p_j^{(\cA)}
=\frac{\partial\psi}{\partial x^{(\cA)\,j}} 
,\ \mbox{and}\ 
z^{(\cA)}
=\psi(\,x^{(\cA)}\,)
\ \right\}.
$$
Thus, one arrives at $\cF_{\psi}=\cA_{\psi}^{\cC}$ in $\cD_{\psi}\subset\cC$. 
In what follows $h_{\psi}=\wh{h}(\Delta_{\psi})$ 
is shown to be a Lyapunov function of 
$\cF_{\psi}$.
The relation 
$$
\wh{h}(\Delta_{\psi})> 0,\quad\mbox{on}\quad \cD_{\psi}\setminus\cA_{\psi}^{\cC},
$$ 
will be used. 
Since 
$$
h_{\psi}|_{\cA_{\psi}^{\cC}}
=\wh{h}_{\psi}(\Delta_{\psi})|_{\Delta_{\psi}=0}
=0,\qquad 
\left.h_{\psi}\right|_{\cD_{\psi}\setminus\cA_{\psi}^{\cC}}
=\wh{h}\,(\Delta_{\psi})\,|_{\cD_{\psi}\setminus\cA_{\psi}^{\cC}}
>0,
$$
and 
$$
\left.\frac{\dr h_{\psi}}{\dr t}\right|_{\cD_{\psi}\setminus\cA_{\psi}^{\cC}}
=\frac{\dr\Delta_{\psi}}{\dr t}\frac{\dr \wh{h}}{\dr\Delta_{\psi}}
\bigg|_{\cD_{\psi}\setminus\cA_{\psi}^{\cC}}
=\bigg(\,\frac{\dr x^j}{\dr t}
\frac{\partial\psi}{\partial x^j}-
\frac{\dr z}{\dr t}
\,\bigg)\,\frac{\dr \wh{h}}{\dr\Delta_{\psi}}\bigg|_{\cD_{\psi}\setminus\cA_{\psi}^{\cC}}
=-\,\wh{h}(\Delta_{\psi})\frac{\dr \wh{h}(\Delta_{\psi})}{\dr\Delta_{\psi}}
\bigg|_{\cD_{\psi}\setminus\cA_{\psi}^{\cC}}
<0,
$$
the function $h_{\psi}$ is a Lyapunov function \cite{HS1974}. 
Thus, according to the stability theorem of Lyapunov, $\cF_{\psi}=\cA_{\psi}^{\cC}$ 
is a set of asymptotically stable fixed points.
\qed
%%%%%%%%%%%5
\end{Proof}
%%%%%%%%%%%
%%%%%%%%%%%%%%%%
\begin{Remark}
%%%%%%%%%%%%%%
Observe that 
$$
\lim_{t\to\infty}\Delta_{\psi}(x(t),z(t))
=0\quad\mbox{and}\quad
\lim_{t\to\infty}\wh{h}\left(\Delta_{\psi}(x(t),z(t))\right)
=0.
$$
%%%%%%%%%%%%%%%
\end{Remark}
%%%%%%%%%%%%%%
%%%%%%%%%%%%%%%%
\begin{Remark}
%%%%%%%%%%%%%%
It follows that 
$$
\left.\Delta_{\psi}\right|_{\Phi_{\cC\cA\psi}\cA_{\psi}}
=0.
$$
%%%%%%%%%%%%%%%
\end{Remark}
%%%%%%%%%%%%%%
%%%%%%%%%%%%%%%%
\begin{Remark}
%%%%%%%%%%%%%%
Physical meaning concerning thermodynamics is as follows. 
Since a Legendre submanifold $\cA_{\psi}^{\cC}$ 
 can physically represent equilibrium 
thermodynamic states, 
Theorem \ref{theorem:main-atrractor-psi-1} can imply that a relaxation process
towards to the equilibrium state is constructed. 
%%%%%%%%%%%%%
\end{Remark}
%%%%%%%%%%%%
%%%%%%%%%%%%%%%%
\begin{Remark}
%%%%%%%%%%%%%%
There exists $n$ invariants in the sense that  
$\dot{x}^j=\cL_{X_h^{\psi}}x^j=0, ( j\in\{1,\ldots,n\} )$.   
%%%%%%%%%%%%%
\end{Remark}
%%%%%%%%%%%%
%%%%%%%%%%
\begin{Remark}
%%%%%%%%
On $\cD_{\psi}$ the contact Hamiltonian $\wh{h}(\Delta_{\psi})$ is of the form 
$$
\wh{h}(\Delta_{\psi})
=\gamma_1\Delta_{\psi}+\gamma_2\Delta_{\psi}^2+\cdots,
$$
with $\gamma_1>0$ and some  $\gamma_2$. 
%%%%%%%%%%%%
\end{Remark}
%%%%%%%%%%%%
%%%%%%%%%%
\begin{Remark}
%%%%%%%%
Choosing $\wh{h}(\Delta_{\psi})=\gamma \Delta_{\psi}$ with $\gamma>0$ in 
Theorem \ref{theorem:main-atrractor-psi-1}, 
one has the  expressions for the contact Hamiltonian flow 
\beq
\dot{x}^j
=0,\qquad
\dot{p}_j
=\gamma\,\left(\,\frac{\partial\, \psi(x)}{\partial x^j}-p_j\,\right),\qquad
\dot{z}
=\gamma\,\left(\,\psi(x)-z\,\right).
\label{contact-Hamiltonian-Lyapunov-psi-vector-exponential-type}
\eeq
whose integral curve passing  a point $(x(0),p(0),z(0))\in\cC$ 
is explicitly expressed as   
\beqa
x^{j}(t)
&=&x^{j}(0),
\non\\ 
p_j(t)
&=&\frac{\partial\,\psi}{\partial x^j}(x(0))
+\left(\,p_j(0)-\frac{\partial\,\psi}{\partial x^j}(x(0))\,\right)\,
\e^{-\,\gamma\, t},
\non\\ 
z(t)
&=&\psi(x(0))+\left(\,z(0)-\psi(\,x(0)\,)
\,\right)\,\e^{-\,\gamma\, t}.
\non
\eeqa
In terms of the introduced 
abbreviations $(x^{\,(\cA)},p^{(\cA)},z^{(\cA)})\in\cA_{\psi}^{\cC}$ the 
above expressions are written as 
$$
x^{j}(t)
=x^{j}(0),\qquad
p_{j}(t)
=p_{j}^{(\,\cA\,)}
+\left(\,
  p_j(0)-p_{j}^{(\,\cA\,)}
\,\right)\,\e^{-\,\gamma\,t},\qquad 
z(t)
=z^{(\,\cA\,)}
+\left(\,
z(0)-z^{(\,\cA\,)}
\,\right)\,\e^{-\,\gamma\,t}.
$$
Notice that the constant 
$\gamma$ can be interpreted as a characteristic time for a relaxation process.
When $\gamma$ is unity, the contact Hamiltonian system  
restricted to a Legendre submanifold 
has briefly been studied in Ref.\cite{Mrugala1991}. 
%%%%%%%%%%
\end{Remark}
%%%%%%%%
%%%%%%%%%%
\begin{Remark}
%%%%%%%%
The idea of this proof can be viewed as a generalized one of the theorem in    
Ref.\cite{Jurkowski2000}.  
%%%%%%%%%%
\end{Remark}
%%%%%%%%

The following is a counterpart of Theorem \ref{theorem:main-atrractor-psi-1}. 
%%%%%%%%%%%%%%
\begin{Thm}
\label{theorem:main-atrractor-varphi-1}
%%%%%%%%%%%%
(Relaxation process in terms of contact Hamiltonian vector field $2$) : 
Let 
$\varphi\in\GamLamC{0}$ be a function of $p$ only,  
and $\cA_{\varphi}$ the Legendre submanifold generated by $\varphi$. 
Define $h_{\varphi},\Delta_{\varphi} \in\GamLamC{0}$, $\cD_{\varphi}\subset\cC$ 
and $\wh{h}:\mathbb{R}\to\mathbb{R}$ 
such that  
\beq
h_{\varphi}(x,p,z)
=\wh{h}(\Delta_{\varphi}),\qquad
\Delta_{\varphi}(x,p,z)
:=x^ip_i-\varphi(p)-z,
\label{contact-Hamiltonian-Lyapunov-varphi}
\eeq
$$
\cD_{\varphi}
:=\left\{\,(x,p,z)\in\cC\ \bigg|\ \wh{h}\,(\Delta_{\varphi})\geq 0,\ 
\mbox{and}\ 
\frac{\dr\, \wh{h}}{\dr \Delta_{\varphi}}>0
 \,\right\}\quad\subset \cC,
$$ 
and
$$
\wh{h}(0)
=0,\qquad\
\wh{h}(\Delta_{\varphi})>0,\ \mbox{for}\ \Delta_{\varphi}\neq 0. 
$$ 
Then, integral curves of the contact vector field associated to $h_{\varphi}$
on $\cD_{\varphi}$ 
connect points of $\cD_{\varphi}$ and those of 
$\Phi_{\cC\cA\varphi}\cA_{\varphi}$. 
Thus, the integral curves can be 
relaxation processes (See Definition \ref{definition-relaxation-process}).
%%%%%%%%%
\end{Thm}
%%%%%%%%%%
%%%%%%%%%%%%%%%
\begin{Proof}
%%%%%%%%%%%%%%
A way to prove this is analogous to the proof of 
Theorem\,\ref{theorem:main-atrractor-psi-1}. 
\qed
%%%%%%%%%
\end{Proof}
%%%%%%%%
%%%%%%%%%%%%%
\begin{Remark}
The contact Hamiltonian vector field $X_h^{\varphi}$
 is the flow that can be expressed by 
\fr{contact-Hamiltonian-vector-components},
 \beq
 X_h^{\varphi}
 =\dot{x}^i\frac{\partial}{\partial x^i}
 +\dot{p}_i\frac{\partial}{\partial p_i}
 +\dot{z}\frac{\partial}{\partial z},
 \ 
 \frac{\dr x^i}{\dr t}
 =\left(\,\frac{\partial\varphi}{\partial p_i}-x^i\,\right)
 \frac{\dr \wh{h}}{\dr \Delta_{\varphi}},
 \ 
 \frac{\dr p_i}{\dr t}
 =0,\ 
 \frac{\dr z}{\dr t}
 =\wh{h}
 +\left(\frac{\partial \varphi}{\partial p_i}-x^i\right)
 p_i\frac{\dr \wh{h}}{\dr\Delta_{\varphi}}.
 \label{contact-Hamiltonian-Lyapunov-varphi-vector}
 \eeq
%%%%%%%
\end{Remark}
%%%%%%%%%%%%

%%%%%%%%%%%%%%%%
\begin{Remark}
%%%%%%%%%%%%%%
Observe that 
$$
\lim_{t\to\infty}\Delta_{\varphi}(x(t),p(t),z(t))
=0\quad\mbox{and}\quad
\lim_{t\to\infty}\wh{h}\left(\Delta_{\varphi}(x(t),p(t),z(t))\right)
=0.
$$
%%%%%%%%%%%%%%%%
\end{Remark}
%%%%%%%%%%%%%%
%%%%%%%%%%%%%%%%
\begin{Remark}
%%%%%%%%%%%%%%
It follows that 
$$
\left.\Delta_{\varphi}\right|_{\Phi_{\cC\cA\varphi}\cA_{\varphi}}
=0.
$$
%%%%%%%%%%%%%%%
\end{Remark}
%%%%%%%%%%%%%%

%%%%%%%%%%%%%%%%
\begin{Remark}
%%%%%%%%%%%%%%
There exists $n$ invariants in the sense that 
$\dot{p}_i=\cL_{X_h^{\varphi}}p_i=0, ( i\in\{1,\ldots,n\} )$.
%%%%%%%%%%%%%
\end{Remark}
%%%%%%%%%%%%
%%%%%%%%%%
\begin{Remark}
%%%%%%%%
On $\cD_{\varphi}$ the contact Hamiltonian $\wh{h}(\Delta_{\varphi})$ 
is of the form 
$$
\wh{h}(\Delta_{\varphi})
=\gamma_1\Delta_{\varphi}+\gamma_2\Delta_{\varphi}^2+\cdots,
$$
with  $\gamma_1>0$ and some $\gamma_2$. 
%%%%%%%%%%%%%
\end{Remark}
%%%%%%%%%%%%
%%%%%%%%%%
\begin{Remark}
%%%%%%%%
Choosing $\wh{h}(\Delta_{\varphi})=\gamma \Delta_{\varphi}$ with $\gamma>0$ in 
Theorem \ref{theorem:main-atrractor-varphi-1}, 
one has the  expressions for the contact Hamiltonian flow 
$$
\dot{x}^i
=\gamma\,\left(\,\frac{\partial\,\varphi(p)}{\partial p_i}-x^i\,\right),
\qquad
\dot{p}_i
=0,\qquad 
\dot{z}
=\gamma\,\left(\,p_i\frac{\partial\,\varphi(p)}{\partial p_i}
-\varphi(p)-z\,
\right),
$$
whose integral curve passing a point $(x(0),p(0),z(0))\in\cC$ 
is explicitly expressed as   
\beqa
x^{i}(t)
&=&\frac{\partial\,\varphi}{\partial p_i}(p(0))
+\left(\,x^i(0)-\frac{\partial\,\varphi}{\partial p_i}(p(0))\,\right)\,\e^{-\,\gamma\, t}, 
\non\\
p_i(t)
&=&p_i(0),
\non\\ 
z(t)
&=&\left(\,p_i(0)\frac{\partial\,\varphi}{\partial p_i}(p(0))-\varphi(p(0))\,\right)
+\left[\,z(0)-\left(\,p_i(0)\frac{\partial\,\varphi}{\partial p_i}(p(0))+\varphi( p(0) )\,\right)
\,\right]\,\e^{-\,\gamma\, t}. 
\non
\eeqa
In terms of 
$(x^{\,(\cA)},p^{(\cA)},z^{(\cA)})\in\cA_{\varphi}^{\cC}:=\Phi_{\cC\cA\varphi}\cA_{\varphi}$ 
the above expressions are written as 
$$
x^{i}(t)
=x^{i\,(\,\cA\,)}+\left(\,
x^i(0)-x^{i\,(\,\cA\,)}
\,\right)\,\e^{-\,\gamma\,t},\qquad
p_{i}(t)
=p_{i}(0),\qquad 
z(t)
=z^{(\,\cA\,)}
+\left(\,
z(0)-z^{(\,\cA\,)}
\,\right)\,\e^{-\,\gamma\,t}.
$$
Notice that the constant 
$\gamma$ can be interpreted as a characteristic time for a relaxation process.
%%%%%%%%%%
\end{Remark}
%%%%%%%%
%%%%%%%%%%
\begin{Remark}
%%%%%%%%
Choosing $\varphi\equiv 0$, and  
$\wh{h}(\Delta_{\varphi})=\gamma \Delta_{\varphi}$ with $\gamma>0$, 
one has the vector field discussed in Ref.\cite{Bravetti-Sep-2014}. 
The authors of that paper 
interpret their particular vector field as   
``near-equilibrium process.''  
In addition, the meaning of their contact Hamiltonian is 
discussed in  Ref.\cite{Bravetti-Sep-2014}. 
%%%%%%%%%%
\end{Remark}
%%%%%%%%

%%%%%%%%%
\subsection{Application of the general theory to a spin system}
\label{subsection-spin-system-relaxation}
%%%%%%%%%
In this subsection, 
attention is concentrated on a 
relaxation dynamics of a spin system whose equilibrium 
state is stipulated in Example\,\ref{example:spin-equilibrium}. 
This is to show how the general theory is applied to physical models. 
As a simple dynamics, the contact Hamiltonian flow 
is taken to be \fr{contact-Hamiltonian-Lyapunov-psi-vector-exponential-type}.
In addition, this contact Hamiltonian system is shown to   
include a kinetic spin model without spin-coupling 
derived with a master equation. 
Although the spin model we employ does not show a phase transition,  
a comparison between the model derived 
in the framework of contact geometry and one derived with a master equation 
is valuable, 
since kinetic spin models have been used for 
elucidating relaxation processes of 
a class of Ising models (See, for example, Ref.\cite{Neda1998}).  
Note that there are many variants of such kinetic models, and the one without 
spin coupling has not explicitly been shown in the literature.  
Therefore, the derivation 
of the kinetic model without spin-coupling is shown in the 
Appendix A of this paper. 

The system is fixed and its macroscopic equilibrium state is defined 
in the following definition. 
%%%%%%%%%%%
\begin{Def}
%%%%%%%%%%%%
(Equilibrium state of a spin system with an external constant magnetic field in contact with a heat bath)  : \ 
Let $\sigma=\pm 1$ be a spin variable, 
$H$ a constant external magnetic field whose dimension is an energy, and  
$\theta=H/(k_{\B}T_{\abs})$. 
Then, the distribution function 
\beq
\mbbP_{\theta}^{\,\can}(\sigma)
=\frac{1}{Z_{\theta}}\exp\left(\,\theta\sigma\,\right), 
\label{canonical-distribution-Ising-model}
\eeq
is referred to as the canonical distribution for the spin system. Here,  
$Z_{\theta}=\exp\left[\,\psi(\theta)\,\right]$ is a partition function 
with $\psi$ being the cumulant generating function : 
\beq
Z_{\theta}
=2\cosh\theta,\qquad 
\psi(\theta)
=\ln Z_{\theta}
=\ln\cosh\theta+\ln 2.
\label{partition-function-canonical-distribution-Ising-model}
\eeq
In addition, the equilibrium value of the magnetization is 
$$
\eta
=\sum_{\sigma=\pm 1}\sigma\,\mbbP_{\theta}^{\,\can}(\sigma)
=\frac{\partial\psi}{\partial\theta}.
$$
The macroscopic equilibrium state for this system is defined to be 
$\cA_{\psi}:=(\theta,\eta(\,\theta\,),\psi(\,\theta))$ with 
$\eta=\partial\psi/\partial\theta$.  
%%%%%%%%%
\end{Def}
%%%%%%%

In the context of nonequilibrium statistical mechanics,
dynamical systems called kinetic Ising models 
have been known and well-studied. 
A simplified kinetic Ising model is the one without spin-coupling.
In the following 
a kinetic spin model without spin-coupling is introduced and that will be 
compared with 
a contact Hamiltonian system. 
%%%%%%%%%
\begin{Def}
%%%%%%%%%%%%
(Kinetic spin model without spin-coupling) : 
The dynamical system 
\beq
\frac{\dr \ave{\sigma}}{\dr t}
=\gamma^{\,\prime}\left(\,\tanh(\beta_{\abs}H)-\ave{\sigma}\,\right)
\label{kinetic-spin-model}
\eeq
is referred to as the kinetic spin model without spin-coupling.  
Here, $\gamma^{\,\prime}$ is a constant,
and $\ave{\sigma}$ is a function of $t$ 
(see the Appendix A for derivation).
%%%%%%%%%
\end{Def}
%%%%%%%%%%

To describe a relaxation process with contact geometry,  
one needs $\psi$ and $(x,p,z)$ 
used in Theorem\,\ref{theorem:main-atrractor-psi-1}.
They are stipulated as follows.
%%%%%%%%
\begin{Postulate}
%%%%%
(Construction of a contact manifold from a given manifold) : 
1. Coordinates $(x,p,z)\in\cC$ of a $3$-dimensional 
contact manifold $(\cC,\lambda)$ can be introduced 
such that $\lambda=\dr z-p\,\dr x$, and that  
$$
\left.(x,p,z)\right|_{\Phi_{\cC\cA\psi}\,\cA_{\psi}}
=(\,\theta,\eta\,(\theta),\psi(\theta)\,).
$$
2. The domain of $\psi$ can be extended such that one can write $\psi(x)$ for 
$\psi\in\GamLamC{0}$. 
%%%%%%%%
\end{Postulate}
%%%%%

Under these postulates one has the following theorem. 
This is the main claim in this subsection, and this supports 
that Theorem\,\ref{theorem:main-atrractor-psi-1}
is a valuable tool for elucidating behaviour of some classes of 
physical models used in 
nonequilibrium thermodynamics. 
%%%%%%%%%%%
\begin{Thm}
%%%%%%%%%%%
(Equivalence of a kinetic model and a contact Hamiltonian system for a spin system) :  
The contact Hamiltonian flow generated by the contact Hamiltonian 
$h_{\psi}=\gamma(\,\psi(x)-p\,)$ is 
\beq
\frac{\dr x}{\dr t}
=0,\qquad
\frac{\dr p}{\dr t}
=\gamma\left(\,\tanh(\beta_{\abs}H)-p\,\right),\qquad
\frac{\dr z}{\dr t}
=\gamma\left(\,\psi- z\,\right).
\label{contact-Hamiltonian-spin-dynamics}
\eeq
In addition, the second equation above is identical to the kinetic 
spin model without spin-coupling, \fr{kinetic-spin-model}.  
%%%%%%%%%%%
\end{Thm}
%%%%%%%%%%
%%%%%%%
\begin{Proof}
%%%%%%%%
Substituting $\psi(x)=\ln\cosh x+\ln 2$ into 
\fr{contact-Hamiltonian-Lyapunov-psi-vector-exponential-type}, 
one has the equations in the theorem. In addition, by identifying
$\ave{\sigma}=p$ and $\gamma^{\,\prime}=\gamma$, one completes the proof. 
\qed
%%%%%%%
\end{Proof}
%%%%%%%%
%%%%
\begin{Remark}
%%%%%%
The last equation of 
\fr{contact-Hamiltonian-spin-dynamics}, involving $\dr z/\dr t$,  
can be interpreted as an equation for the time-evolution 
of the dimensionless negative 
Helmholtz free energy.
%%%%
\end{Remark}
%%%%%%

%%%%%%%%%%%%%%%%
\section{Characterization of relaxation process in terms of the 
Mrugala  metric tensor field}
\label{sec-calculations-with-metric}
%%%%%%%%%%%%%%
Once a metric tensor field on a manifold is introduced,   
one can retrieve more information about vector fields.
In this section, it is shown 
some relations between the contact Hamiltonian vector fields 
discussed in the previous section 
and tangent vector fields on Legendre submanifolds. 

%%%%%%%%%%%5
\subsection{Mrugala metric tensor field}
%%%%%%%%%%%\
In this subsection, a well-known metric tensor field is introduced 
and its basic mathematical features are summarized. 
 
The following metric tensor field on a contact manifold  has been often used
in the context of 
geometric thermodynamics.
%%%%%%%%%
\begin{Def}
%%%%%%%% 
(Mrugala metric tensor field, \cite{MrugalaX}) :  
The metric tensor field $G\in\GtC{0}{2}$ 
\beq
G=\dr x^a\sympro\dr p_a
+\lambda\otimes\lambda,\quad
\quad\mbox{where}\quad 
\dr x^a\sympro\dr p_a
:=\frac{1}{2}\dr x^a\otimes \dr p_a+\frac{1}{2}\dr p_a\otimes \dr x^a, 
\label{def-Mrugala-metric}
\eeq
is referred to as the Mrugala metric tensor field. 
%%%%%%%%%%
\end{Def}
%%%%%%%%
%%%%%%%%%%%%%%
\begin{Remark}
%%%%%%%%%%%%%%
This metric tensor field is pseudo-Riemannian \cite{Preston2008}. 
%%%%%%%%%%%%%%
\end{Remark}
%%%%%%%%%%%%%%
%%%%%%%%%%%%%%
\begin{Remark}
%%%%%%%%%%%%%%
The factor $1/2$ in \fr{def-Mrugala-metric} 
is important when comparing the Fisher metric tensor field  
used in information geometry.  
%%%%%%%%%%
\end{Remark}
%%%%%%%%%%%%%
Detailed studies on this metric tensor field 
are found in Refs.\cite{Preston2008}, \cite{Mrugala2005}, and
\cite{Bravetti-Aug-2014}. 

%%%%%%% 
\begin{Thm}
  (Killing vector fields, \cite{Preston2008}) : 
%%%%%%%
Let 
$G$ be the Mrugala metric tensor field. 
Then, the vector fields  
$$
Q_l^k
:=p_l\frac{\partial}{\partial p_k}-x^k\frac{\partial}{\partial x^l},\qquad
R=\frac{\partial}{\partial z},\qquad
A^a
:=\frac{\partial}{\partial p_a}+x^a\frac{\partial}{\partial z},\qquad
B_a
:=-\,\frac{\partial}{\partial x^a},\qquad a,l,k\in\{1,\ldots,n\},
$$
are Killing. 
In addition, these vector fields are expressed as contact Hamiltonian 
vector fields with 
$$
h_{Q_l^k}
=x^kp_l,\quad
h_R
=1,\quad
h_{A^a}
=x^a,\quad
h_{B_a}
=p_a.
$$
%%%%%%% 
\end{Thm}
%%%%%%%
%%%%%%
\begin{Proof}
%%%%%
It follows from straightforward calculations that 
$\cL_{Q_l^k}G=0$, $\cL_RG=0$, $\cL_{A^a}G=0$, and $\cL_{B_a}G=0$
for all $a,l,k$. In addition, substituting the assumed contact Hamiltonians 
into  \fr{contact-Hamiltonian-vector-components}, one can verify that 
the Killing vector fields are the contact Hamiltonian vector fields. 
\qed
%%%%%%
\end{Proof}
%%%%%
%%%%%%%%%%%%5
\begin{Def}
%%%%%%%%%%%%
(Metric dual) : 
Let 
$Z$ be an arbitrary vector field. Then, the one-form  
$$
Z^{\sharp}
:=G(Z,-)
=G(-,Z), \quad\in\GamLamC{1}
$$ 
is referred to as the metric dual of $Z$.
%%%%%%%%%
\end{Def}
%%%%%%%%%%%%%
%%%%%%%
\begin{Thm}
%%%%%%%
  (Geodesics, \cite{Preston2008}) : 
Let 
$G$ be the 
Mrugala metric tensor field, and 
$\nabla$ the Levi-Civita connection derived from $G$.  
Then, the following vector fields: 
$$
R=\frac{\partial}{\partial z},\qquad 
P^a:=\frac{\partial}{\partial p_a},\qquad 
L_a:=\frac{\partial}{\partial x^a}+p_a\frac{\partial}{\partial z},
\qquad a\in\{1,\ldots,n\},
$$ 
give geodesics. 
%%%%%%% 
\end{Thm}
%%%%%%%
%%%%%%%%
\begin{Proof}
%%%%%%%%
In what follows 
$\nabla_RR=\nabla_{P^a}P^a=\nabla_{L_a}L_a=0, (a\in\{1,\ldots,n\})$, 
(no sum) are proved with the formula 
\beq
\nabla_XX^{\sharp}
=\cL_{X}X^{\sharp}-\frac{1}{2}\,\dr\left(G(X,X)\right),
\label{formula-Levi-Civita-vector}
\eeq 
for arbitrary $X\in\GTC$.
To prove $\nabla_RR=0$, 
it follows from 
\beq
R^{\sharp}
=G(R,-)
=\lambda \in\GamLamC{1}\quad \mbox{and}\quad
G(R,R)
=\lambda(R)
=1,
\label{Reeb-vector-metric-dual-normalize}
\eeq  
that 
$$
\nabla_{R}R^{\sharp}
=\cL_{R}R^{\sharp}-\frac{1}{2}\dr\,\left[\,G(R,R)\,\right]
=\cL_{R}\lambda
=0, 
$$ 
where we have used \fr{remark-Reeb-vector-Lie-derivative}. 
With this equation and the property that the Levi-Civita connection 
is a type of metric-compatible connection, 
one concludes that 
\beq
\nabla_RR
=0,
\label{Reeb-vector-metric-geodesic}
\eeq
from which $R$ gives geodesics. 
To prove $\nabla_{P^a}P^a=0,$ (no sum), it follows from  
$P^{a\,\sharp}=G(P^a,-)=(1/2)\,\dr x^a$ 
that $\cL_{P^a}P^{a\,\sharp}=0$. Substituting $\cL_{P^a}P^{a\,\sharp}=0$ and 
$G(P^a,P^a)=P^{a\,\sharp}\,(P^a)=0,$ (no sum) 
into \fr{formula-Levi-Civita-vector}, one has 
$\nabla_{P^a}P^{a\,\sharp}=0$. This yields $\nabla_{P^a}P^{a}=0$.    
To prove $\nabla_{L_a}L_a=0,$ (no sum), it follows from  
$L_a^{\,\sharp}=G(L_a,-)=(1/2)\,\dr p_a$  
that $\cL_{L_a}L_{a}^{\,\sharp}=0$. Substituting $\cL_{L_a}L_{a}^{\,\sharp}=0$ and 
$G(L_a,L_a)=L_a^{\,\sharp}\,(L_a)=0,$ (no sum) 
into \fr{formula-Levi-Civita-vector}, one has 
$\nabla_{L_a}L_{a}^{\,\sharp}=0$. This yields $\nabla_{L_a}L_{a}=0$.    
\qed
%%%%%%%
\end{Proof}
%%%%%%
%%%%%%%%
\begin{Remark}
%%%%%%%5
It follows that $R\in\ker(\dr\lambda)$ and $\{P^a\},\{L_a\}\in\ker(\lambda)$, 
where 
$$
\ker(\beta)
:=\left\{\,Z\in\GTC\,|\,\ii_Z\beta=0\,\right\},
$$
for some $\beta\in\GamLamC{q},(q\in\{1,\ldots,n\})$.
%%%%%%%%
\end{Remark}
%%%%%%%5

One of the reasons why the Mrugala metric tensor field is often used 
in the literature is the following. 
%%%%%%%%%
\begin{Thm}
\label{theorem-Legendre-submanifold-is-statistical-manifold-Mrugala}
%%%%%%% 
(Pull-back of the Mrugala metric tensor field on a Legendre submanifold) : 
Let 
$\psi\in\GamLamC{0}$ be a strictly 
convex function of $x$ only, 
$\varphi\in\GamLamC{0}$ a strictly 
convex function of $p$ only,  
and $G$ the Mrugala metric tensor field \fr{def-Mrugala-metric}. 
Then, 
$(\cA_{\psi},\Phi_{\cC\cA\psi}^{\ \ \ \ *}G)$ and  
$(\cA_{\varphi},\Phi_{\cC\cA\varphi}^{\ \ \ \ *}G)$
are identical to the $n$-dimensional 
Riemannian manifolds $(\cA_{\psi},g^{\,\cA\psi})$ and 
$(\cA_{\varphi},g^{\,\cA\varphi})$ given in 
Theorem \ref{theorem-contact-manifold-convex-function-induce-Riemannian}, 
respectively. 
%%%%%%%%%
\end{Thm}
%%%%%%%
%%%%%%%%
\begin{Proof}
%%%%%%%
It follows that 
$$
\Phi_{\cC\cA\psi}^{\ \ \ \ *}G
=\Phi_{\cC\cA\psi}^{\ \ \ \ *}\left(\,\dr x^a\sympro\dr p_a\,\right)
=\Phi_{\cC\cA\psi}^{\ \ \ \ *}\left(\,
\frac{\partial^2\psi}{\partial x^a\partial x^b}\,\right)\,\dr \theta^a
\otimes\dr \theta^b
=g^{\,\cA\psi},\qquad 
\theta^a
:=\Phi_{\cC\cA\psi}^{\ \ \ \ *} x^a,
$$
and 
$$
\Phi_{\cC\cA\varphi}^{\ \ \ \ *}G
=\Phi_{\cC\cA\varphi}^{\ \ \ \ *}\left(\,\dr x^a\sympro\dr p_a\,\right)
=\Phi_{\cC\cA\varphi}^{\ \ \ \ *}\left(\,
\frac{\partial^2\psi}{\partial p_a\partial p_b}\,\right)
\,\dr \eta_a\otimes\dr \eta_b
=g^{\,\cA\varphi},\qquad 
\eta_a
:=\Phi_{\cC\cA\varphi}^{\ \ \ \ *} p_a.
$$
\qed
%%%%%%%%
\end{Proof}
%%%%%%%

%%%%%%%%%%%%%
\subsection{Relations between lower dimensional manifolds and 
contact Hamiltonian functions}
%%%%%%%%%%%%
In this subsection,   
{\it control manifold}, lower-dimensional manifold in a contact manifold 
introduced in Ref.\cite{Bravetti-Aug-2014}, is defined. Then,  
the characterization of the pull-back of $\Delta_{\psi}$ 
in Theorem \ref{theorem:main-atrractor-psi-1}
and that of 
$\Delta_{\varphi}$ in Theorem \ref{theorem:main-atrractor-varphi-1} are given.  
%%%%%%%%%%
\begin{Def}
\label{def-control-manifold}
%%%%%%%%%%
(Control manifold, \cite{Bravetti-Aug-2014}) : 
Let 
$\psi\in\GamLamC{0}$ be a strictly convex 
function of $x$ only, and   
$\varphi\in\GamLamC{0}$ a strictly convex 
function of $p$ only. 
Define $(n+1)$-dimensional manifolds  
$\cB_{\psi}$ and $\cB_{\varphi}$ as  
\beq
\Phi_{\cC\cB\psi}\cB_{\psi}
=\left\{\ (x,p,z)\in\cC\ \bigg|\ p_j=\frac{\partial\psi}{\partial x^j},\quad 
  j\in \{1,\ldots,n\}\right\}
\label{control-manifold-psi}
\eeq
and 
\beq
\Phi_{\cC\cB\varphi}\cB_{\varphi}
=\left\{\ (x,p,z)\in\cC\ \bigg|\ x^i=\frac{\partial\varphi}{\partial p_i},\quad 
  i\in \{1,\ldots,n\}\right\},
\label{control-manifold-varphi}
\eeq
respectively, where $\Phi_{\cC\cB\psi}:\cB_{\psi}\to\cC$ and 
$\Phi_{\cC\cB\varphi}:\cB_{\varphi}\to\cC$ are embeddings. 
They are referred to as control manifolds.
%%%%%%%%%%
\end{Def}
%%%%%%%%%%
%%%%%%%%%
\begin{Postulate}
%%%%%%%%%
In this paper, we assume that $\Phi_{\cC\cB\psi}\,\cB_{\psi}$ and 
$\Phi_{\cC\cB\varphi}\,\cB_{\varphi}$ 
are submanifolds of a contact manifold $(\cC,\lambda)$.
%%%%%%%%%
\end{Postulate}
%%%%%%%%%

On $\cB_{\psi}$ and $\cB_{\varphi}$, the induced  
metric tensor fields and vector fields giving geodesics  
are calculated as follows. 
%%%%%%%%%%
\begin{Lemma}
%%%%%%%%%%
\label{lemma-induced-metric-geodesic-contact-manifold}
Let 
$\psi\in\GamLamC{0}$ be a strictly 
convex function of $x$ only,  
$\varphi\in\GamLamC{0}$ a strictly  
convex function of $p$ only,  
$\Phi_{\cC\cB\psi}\,\cB_{\psi}$ and 
$\Phi_{\cC\cB\varphi}\,\cB_{\varphi}$  
the $(n+1)$-dimensional submanifolds defined by  
\fr{control-manifold-psi} and \fr{control-manifold-varphi}, respectively, 
and 
$G$ the Mrugala metric tensor field 
\fr{def-Mrugala-metric}. Then, the induced metric tensor fields  
$G^{\cB\psi}:=\Phi_{\cC\cB\psi}^{\ \ \ \ *}G$ and
$G^{\cB\varphi}:=\Phi_{\cC\cB\varphi}^{\ \ \ \ *}G$     
are calculated to be 
\beq
G^{\cB\psi}
=\frac{\partial^2\psi}{\partial x^a\partial x^b}\dr x^a\otimes\dr x^b
+\lambda^{\,\cB\psi}\otimes
\lambda^{\,\cB\psi}
\label{induced-metric-contact-psi}
\eeq
and 
\beq
G^{\cB\varphi}
=\frac{\partial^2\varphi}{\partial p_a\partial p_b}\dr p_a\otimes\dr p_b
+\lambda^{\,\cB\varphi}\otimes
\lambda^{\,\cB\varphi},
\label{induced-metric-contact-varphi}
\eeq
where 
\beq
\lambda^{\,\cB\psi}
:=\Phi_{\cC\cB\psi}^{\ \ \ \ *}\lambda
=\dr z-\frac{\partial \psi}{\partial x^j}\dr x^j
=\dr \left(z-\psi(x)\right)\quad\mbox{and}\quad
\lambda^{\,\cB\varphi}
:=\Phi_{\cC\cB\varphi}^{\ \ \ \ *}\lambda
=\dr z-p_i\frac{\partial^2\varphi}{\partial p_i\partial p_b}\dr p_b.
\label{induced-contact-form-contact-manifold}
\eeq
In addition, it follows that 
$$
\nabla^{\cB\psi}_{R_\psi}R_{\psi}
=0\qquad\mbox{and}\qquad 
\nabla^{\cB\varphi}_{R_{\varphi}}R_{\varphi}
=0,
$$
where 
$R_{\psi}:=\partial/\partial z\in\GTB_{\psi}$, 
$R_{\varphi}:=\partial/\partial z\in\GTB_{\varphi}$, 
$\nabla^{\cB\psi}$ and $\nabla^{\cB\varphi}$ are the Levi-Civita connections 
uniquely determined by $G^{\cB\psi}$ and $G^{\cB\varphi}$, respectively.    
%%%%%%%%%%
\end{Lemma}
%%%%%%%%%%
%%%%%%%%%%
\begin{Proof}
%%%%%%%%%%
It is straightforward to show the explicit forms of 
$G^{\cB\psi},G^{\cB\varphi},\lambda^{\,\cB\psi}$, and $\lambda^{\,\cB\varphi}$. To show 
$\nabla^{\cB\psi}_{R_\psi}R_{\psi}=0,$ and $\nabla^{\cB\varphi}_{R_{\varphi}}R_{\varphi}=0$,
one uses \fr{formula-Levi-Civita-vector}, and the fact 
that the Levi-Civita connection is a type of metric compatible connection.
\qed
%%%%%%%%%%
\end{Proof}
%%%%%%%%%%

On control manifolds and Legendre submanifolds embedded in a contact manifold,  
one has the following theorem.
%%%%%%%%%
\begin{Thm}
%%%%%%%%%
Let 
$\psi\in\GamLamC{0}$ be a strictly 
convex function of $x$ only, 
$\varphi\in\GamLamC{0}$ a strictly 
convex function of $p$ only, 
$\cA_{\psi}$ and $\cA_{\varphi}$ the Legendre submanifolds generated by $\psi$ and 
$\varphi$, respectively, 
$\Phi_{\cC\cA\psi}:\cA_{\psi}\to\cC,
\Phi_{\cC\cA\varphi}:\cA_{\varphi}\to\cC,
\Phi_{\cC\cB\psi}:\cB_{\psi}\to\cC,
\Phi_{\cC\cB\varphi}:\cB_{\varphi}\to\cC,
\Phi_{\cB\cA\psi}:\cA_{\psi}\to\cB_{\psi},
\Phi_{\cB\cA\varphi}:\cA_{\varphi}\to\cB_{\varphi},
$ embeddings (See the diagram below), and  
$G,\cB_{\psi},\cB_{\varphi},G^{\,\cB\psi},G^{\,\cB\varphi},\lambda^{\,\cB\psi},\lambda^{\,\cB\varphi}$ given by 
\fr{def-Mrugala-metric}, 
\fr{control-manifold-psi}, 
\fr{control-manifold-varphi}, 
\fr{induced-metric-contact-psi}, \fr{induced-metric-contact-varphi},
\fr{induced-contact-form-contact-manifold}.
 $$
 \xymatrix{
\cB_{\psi}
 \ar[rrd]^{\Phi_{\cC\cB\psi}}
& &&&\cB_{\varphi}
 \ar[lld]_{\Phi_{\cC\cB\varphi}}\\
& &\cC& &\\
\cA_{\psi}
 \ar[rru]_{\Phi_{\cC\cA\psi}}
 \ar[uu]^{\Phi_{\cB\cA\psi}}
& && &\cA_{\varphi}
 \ar[llu]^{\Phi_{\cC\cA\varphi}}
 \ar[uu]_{\Phi_{\cB\cA\varphi}}
 } 
 $$
Then, the following diagram commutes: 
$$
\xymatrix{
(\lambda^{\cB\psi},G^{\,\cB\psi})
\ar[dd]_{\Phi_{\cB\cA\psi}^{\ \ \ \ *}}
&&(\lambda^{\cB\varphi},G^{\,\cB\varphi})
\ar[dd]^{\Phi_{\cB\cA\varphi}^{\ \ \ \ *}}
\\
&(\lambda,G)
\ar[lu]_{\Phi_{\cC\cB\psi}^{\ \ \ \ *}}
\ar[ru]^{\Phi_{\cC\cB\varphi}^{\ \ \ \ *}}
\ar[ld]^{\Phi_{\cC\cA\psi}^{\ \ \ \ *}}
\ar[rd]_{\Phi_{\cC\cA\varphi}^{\ \ \ \ *}}
& \\
(0,g^{\,\cA\psi})
&&(0,g^{\,\cA\varphi})
} 
$$
%%%%%%%%%
\end{Thm}
%%%%%%%%%

%%%%%%%%%%
\begin{Proof}
%%%%%%%%%%
 It follows from 
Theorem \ref{theorem-Legendre-submanifold-is-statistical-manifold-Mrugala} 
that the diagrams relating  
$(0,g^{\,\cA\psi})$, $(0,g^{\,\cA\varphi})$, 
and $(\lambda,G)$ hold. 
For the diagram between $(\lambda^{\,\cB\psi},G^{\,\cB\psi})$ 
and $(0,g^{\,\cA\psi})$, and that between 
$(\lambda^{\,\cB\varphi},G^{\,\cB\varphi})$ 
and $(0,g^{\,\cA\varphi})$,  
it follows from  
$\Phi_{\cB\cA\psi}^{\ \ \ \ *}\lambda^{\,\cB\psi}=0$,  
$\Phi_{\cB\cA\varphi}^{\ \ \ \ *}\lambda^{\,\cB\varphi}=0$, and 
Lemma \ref{lemma-induced-metric-geodesic-contact-manifold}
that 
$$
\Phi_{\cB\cA\psi}^{\ \ \ \ *}G^{\cB\psi}
=\Phi_{\cB\cA\psi}^{\ \ \ \ *}\left(\frac{\partial^2\psi}{\partial x^a\partial x^b}
\right)\dr \theta^a\otimes\dr \theta^b
=g^{\,\cA\psi},\qquad 
\theta^a
:=\Phi_{\cC\cA\psi}^{\ \ \ \ *}x^a,
$$
and 
$$
\Phi_{\cB\cA\varphi}^{\ \ \ \ *}G^{\cB\varphi}
=\Phi_{\cB\cA\varphi}^{\ \ \ \ *}\left(\frac{\partial^2\varphi}{\partial p_a\partial p_b}
\right)\dr \eta_a\otimes\dr \eta_b
=g^{\,\cA\varphi},\qquad 
\eta_a
:=\Phi_{\cC\cA\varphi}^{\ \ \ \ *}p_a.
$$
\qed
%%%%%%%%%%
\end{Proof}
%%%%%%%%%%

Then, one has the following theorem for providing an equation 
for the pull-back of an element of the contact 
Hamiltonian given in Theorem \ref{theorem:main-atrractor-psi-1}.  
%%%%%%%%%%%%%%%%%
\begin{Thm}
\label{theorem-harmonic-psi}
%%%%%%%%%%%%%%%%%
(Harmonic function on control manifold $1$) : 
Let 
$\psi\in\GamLamC{0}$ be a strictly convex function of $x$ only,  
$\cB_{\psi}$ the control manifold defined in  
\fr{control-manifold-psi}, 
and $G$ the Mrugala metric tensor field 
\fr{def-Mrugala-metric}.
Define $\Delta_{\cB\psi} \in\GamLamB{0}_{\psi}$ to be 
$\Delta_{\cB\psi}:=\psi(x)-z=\Phi_{\cC\cB\psi}^{\ \ \ \ *}\Delta_{\psi}$
with $\Delta_{\psi}\in\GamLamC{0}$ defined in Theorem
\ref{theorem:main-atrractor-psi-1}.
Then, $\Delta_{\cB\psi}$ is a harmonic function on 
$\cB_{\psi}$ :
$$
\star_{\psi}^{-1}\dr \star_{\psi}\dr \Delta_{\cB\psi}
=0,
$$
where 
$\star_{\psi}:\GamLamB{q}_{\psi}\to\GamLamB{n+1-q}_{\psi}, (q\in\{0,\ldots,n+1\})$  
is the Hodge dual map 
with $\star_{\psi} 1$ 
being the canonical volume form  
on the (pseudo-) Riemannian manifold $(\cB_{\psi},\Phi_{\cC\cB\psi}^{\ \ \ \ *}G)$, 
and $\star_{\psi}^{-1}$ the inverse map of $\star_{\psi}$.
%%%%%%%%%%%%%%%%%
\end{Thm}
%%%%%%%%%%%%%%%%%
%%%%%%%%
\begin{Proof}
%%%%%%%%
To prove this, one uses the following statement.
Let $(\cM,g)$ be a Riemannian or pseudo-Riemannian manifold, and $\star$  
the Hodge dual map. 
If $K\in\GTM$ is Killing, and $f\in\GamLamM{0}$ is such that $\dr f=g(K,-)$,
then 
\beq
\star^{-1}\dr\star\dr f
=0
\label{Laplacian-and-Killing-formula}
\eeq
( see the Appendix B for a proof ).

Define 
$$
\lambda^{\cB\psi}
:=\Phi_{\cC\cB\psi}^{\ \ \ \ *}\lambda\quad
\mbox{ and }\quad
G^{\cB\psi}
:=\Phi_{\cC\cB\psi}^{\ \ \ \ *}G,
$$
whose local expressions have been given in 
\fr{induced-contact-form-contact-manifold} 
and \fr{induced-metric-contact-psi}.  
Observe that $R_{\psi}=\partial/\partial z\in\GTB_{\psi}$ 
is a Killing vector field,  
$$
\cL_{R_{\psi}}G^{\cB\psi}
=0,
$$
and that 
$$
G^{\cB\psi}(R_{\psi},-)
=\lambda^{\cB\psi}
=\dr (z-\psi)
=-\,\dr \Delta_{\cB\psi}.
$$
Identifying $\cM=\cB_{\psi}$, $g=G^{\cB\psi}$, $K=R_{\psi}$, $f=-\Delta_{\cB\psi}$ and 
substituting these into 
\fr{Laplacian-and-Killing-formula}, one completes the proof. 
\qed
%%%%%%%%
\end{Proof}
%%%%%%%%
%%%%%%%%%%
\begin{Remark}
%%%%%%%%%%
It is straightforward to show that 
$$
\star_{\psi}^{-1}\dr\star_{\psi}\dr\,\bigg(\,
\underbrace{\cL_{R_{\psi}}\cL_{R_{\psi}}\cdots\cL_{R_{\psi}}}_{q}\,\Delta_{\cB\psi}\,\bigg)
=0.\qquad q=0,1,2,\ldots.
$$
%%%%%%%%%%
\end{Remark}
%%%%%%%%%%

The following is a counterpart of this theorem. 
%%%%%%%%%%%%%%%%%
\begin{Thm}
\label{theorem-harmonic-varphi}
%%%%%%%%%%%%%%%%%
(Harmonic function on control manifold $2$) : 
Let 
$\varphi\in\GamLamC{0}$ be a strictly convex function of $p$ only,  
$\cB_{\varphi}$ the control manifold  defined in  
\fr{control-manifold-varphi}, 
and $G$ the Mrugala metric tensor field 
\fr{def-Mrugala-metric}.
Define $\Delta_{\cB\varphi} \in\GamLamB{0}_{\varphi}$ to be 
$\Delta_{\cB\varphi}:=x^ip_i-z-\varphi(p)
=\Phi_{\cC\cB\varphi}^{\ \ \ \ *}\Delta_{\varphi}$
with $\Delta_{\varphi}\in\GamLamC{0}$ defined in Theorem
\ref{theorem:main-atrractor-varphi-1}.
Then, $\Delta_{\cB\varphi}$ is a harmonic function on 
$\cB_{\varphi}$, 
$$
\star_{\varphi}^{-1}\dr \star_{\varphi}\dr \Delta_{\cB\varphi}
=0,
$$
where $\star_{\varphi}:\GamLamB{q}_{\varphi}\to\GamLamB{n+1-q}_{\varphi}, (q\in\{0,\ldots,n+1\})$ 
is the Hodge dual map 
with $\star_{\varphi} 1$ 
being the canonical volume form on 
the (pseudo-) Riemannian manifold $(\cB_{\varphi},\Phi_{\cC\cB\varphi}^{\ \ \ \ *}G)$,
and $\star_{\varphi}^{-1}$ the inverse map of $\star_{\varphi}$. 
%%%%%%%%%%%%%%%%%
\end{Thm}
%%%%%%%%%%%%%%%%%
%%%%%%%%
\begin{Proof}
%%%%%%%%
A way to prove this is analogous to the proof  
of Theorem\,\ref{theorem-harmonic-psi}.
\qed
%%%%%%%%
\end{Proof}
%%%%%%%%
%%%%%%%%%%
\begin{Remark}
%%%%%%%%%%
It is straightforward to show that 
$$
\star_{\varphi}^{-1}\dr\star_{\varphi}\dr\,\bigg(\,
\underbrace{\cL_{R_{\varphi}}\cL_{R_{\varphi}}\cdots\cL_{R_{\varphi}}}_{q}\,\Delta_{\cB\varphi}\,\bigg)
=0.\qquad q=0,1,2,\ldots.
$$
%%%%%%%%%%
\end{Remark}
%%%%%%%%%%

A physical meaning of Theorem 
\ref{theorem-harmonic-psi} 
and that of Theorem \ref{theorem-harmonic-varphi}
are not known so far. However, since the Laplace equation plays a role in 
physics in general, it is expected that implications of these theorems 
will be found in the study of nonequilibrium thermodynamics.

%%%%%%%%%%%%%%%%%%%%%%%%%%%%%%%%%%%%%%%%%%%%%%%%%%%%%%%%%%%%%%
\subsection{Tangent vector fields of Legendre submanifolds}
%%%%%%%%%%%%%%%%%%%%%%%%%%%%%%%%%%%%%%%%%%%%%%%%%%%%%%%%%%%%5
Tangent vector fields on Legendre submanifolds are calculated in this subsection. These vector fields physically represent quasi-stationary processes.   
The resultant calculations here will be used in the next subsection to 
characterize the relaxation processes.

%%%%%%%%
\begin{Proposition}
\label{theorem:tangent-vector-Legendre-submanifold-psi}
%%%%%%%\
(Tangent vector field of Legendre submanifold $1$) : 
Let 
$\psi\in\GamLamC{0}$ be a strictly convex function of $x$ only.
Define the tangent space 
$T_{x}(\Phi_{\cC\cA\psi}\,\cA_{\psi}), (\, x\in\Phi_{\cC\cA\psi}\,\cA_{\psi}\, )$ 
as 
$$
T_{x}(\Phi_{\cC\cA\psi}\,\cA_{\psi})
=\left\{\ Y\in T_{\xi}\cC\ |\ 
\dr\Psi_0^{\psi}(Y)
=0\ \mbox{and}\quad 
\dr\Psi_j^{\psi}(Y)
=0,\,j\in\{1,\ldots,n\}\,\right\},
\qquad x\in\Phi_{\cC\cA\psi}\,\cA_{\psi}
$$
with $\Psi_0^{\psi}:\cC\to\mathbb{R}$ and 
$\Psi_j^{\psi}:\cC\to\mathbb{R}$, being 
$$
\Psi_0^{\psi}(x,p,z)
:=z-\psi(x),\qquad 
\Psi_j^{\psi}(x,p,z)
:=p_j-\frac{\partial \psi}{\partial x^j},\quad j\in\{1,\ldots,n\}.
$$
Every vector field 
$Y\in T_{x}(\Phi_{\cC\cA\psi}\,\cA_{\psi})$  
is then of the form
\beq
Y=\dot{x}^jY_j^{\psi},\quad 
Y_j^{\psi}
:=\frac{\partial}{\partial x^j}
+\frac{\partial^2\psi}{\partial x^j\partial x^b}\frac{\partial}{\partial p_b}
+\frac{\partial\psi}{\partial x^j}\frac{\partial}{\partial z},\qquad 
\dot{x}^j
=\dot{x}^j(x,p,z).
\label{Legendre-submanifold-tangent-vector-psi}
\eeq
%%%%%
\end{Proposition}
%%%%%
%%%%%%%%
\begin{Proof} 
%%%%%%%%%%%
In general $Y\in T_{\xi}\cC$ can be written as 
$$
Y=\dot{x}^j\frac{\partial}{\partial x^j}
+\dot{p}_j\frac{\partial}{\partial p_j}+\dot{z}\frac{\partial}{\partial z}.
$$
Substituting this form into the $(n+1)$ conditions, one has
constraints for $\dot{x}^j,\dot{p}_j$ and $\dot{z}$,   
\beqa
0&=&Y\Psi_0^{\psi}
=Y(z-\psi(x))
=\dot{z}-\dot{x}^j\frac{\partial \psi(x)}{\partial x^j},
\non\\
0&=&Y\Psi_j^{\psi}
=Y\left(p_j-\frac{\partial\psi(x)}{\partial x^j}\right)
=\dot{p}_j-\dot{x}^b\frac{\partial^2\psi(x)}{\partial x^j\partial x^b}, 
\non
\eeqa
where we have used \fr{example-psi-Legendre-submanifold}.
Thus, $\dot{z}$ and $\{\dot{p}_j\}$ are written in terms of $\{\dot{x}^j\}$. 
Substituting these equations, one has
$$
Y=\dot{x}^j\left[\,\frac{\partial}{\partial x^j}
+\frac{\partial^2\psi}{\partial x^j\partial x^b}\frac{\partial}{\partial p_b}
+\frac{\partial\psi}{\partial x^j}\frac{\partial}{\partial z}\,\right],\qquad 
\dot{x}^j
=\dot{x}^j(x,p,z).
$$
\qed
%%%%%
\end{Proof}
%%%%%
%%%%%%%
\begin{Remark}
%%%%%%%
It follows that 
$$
\Phi_{\cC\cA\psi}^{\ \ \ \ *}\left[\,\lambda(Y_j^{\psi})\,\right]
=0\quad\mbox{and}\quad
\lambda\left(\left.Y_j^{\psi}\right|_{\Phi_{\cC\cA\psi}\cA_{\psi}}\right)
=0,\quad
\qquad j\in\{1,\ldots,n\}.
$$
%%%%%%%
\end{Remark}
%%%%%%%
%%%%%%%
\begin{Remark}
%%%%%%%
The expression for $T_{x}(\Phi_{\cC\cA\psi}\,\cA_{\psi})$ 
may be equivalent to the one in discussed in Ref.\cite{GC2011},  
$$
T_{x}(\Phi_{\cC\cA\psi}\,\cA_{\psi})
=\left\{\,X\in T_{\xi}\cC\, |\, 
\ii_X\left(\dr\Psi_0\wedge\dr\Psi_1\wedge\cdots\wedge\dr\Psi_n\right)=0
\,\right\}.
$$  
%%%%%%%
\end{Remark}
%%%%%%%

The following is a counterpart of this proposition. 
%%%%%%%%%%
\begin{Proposition}
%%%%%%%%%
(Tangent vector field of Legendre submanifold $2$) :  
Let 
$\varphi\in\GamLamC{0}$ be a strictly convex function of $p$ only.
Define the tangent space 
$T_{p}(\Phi_{\cC\cA\varphi}\,\cA_{\varphi}), ( p\in\Phi_{\cC\cA\varphi}\,\cA_{\varphi})$ 
as   
$$
T_{p}(\Phi_{\cC\cA\varphi}\,\cA_{\varphi})
=\left\{\ Y\in T_{\xi}\cC\ |\ 
\dr\Psi_{\varphi}^{0}(Y)
=0\ \mbox{and}\quad 
\dr\Psi_{\varphi}^i(Y)
=0,\,i\in\{1,\ldots,n\}\ \right\},
\qquad p\in\Phi_{\cC\cA\varphi}\,\cA_{\varphi}
$$
with $\Phi_{\varphi}^0:\cC\to\mathbb{R}$ and 
$\Phi_{\varphi}^i:\cC\to\mathbb{R}$,  being 
$$
\Psi_{\varphi}^0(x,p,z)
=z-\left(\,p_i\frac{\partial\varphi}{\partial p_i}-\varphi(p)\right),\qquad 
\Psi_{\varphi}^i(x,p,z)
=x^i-\frac{\partial \varphi}{\partial p_i},\quad i\in\{1,\ldots,n\}.
$$
Every vector field $Y\in T_{p}(\Phi_{\cC\cA\varphi}\,\cA_{\varphi})$ 
 is then of the form
\beq
Y=\dot{p}_iY_{\varphi}^i,\quad 
Y_{\varphi}^i
:=\frac{\partial^2\varphi}{\partial p_i\partial p_b}
\frac{\partial}{\partial x^b}
+\frac{\partial}{\partial p_i}
+p_b\frac{\partial^2\varphi}{\partial p_i\partial p_b}\frac{\partial}{\partial z},\qquad 
\dot{p}_i
=\dot{p}_i(x,p,z).
\label{Legendre-submanifold-tangent-vector-varphi}
\eeq
%%%%%%%%%
\end{Proposition}
%%%%%%%%
%%%%%%%%%%%%%%
\begin{Proof}
%%%%%%%%%%%%%
A way to prove this is analogous to the proof of Proposition\,
\ref{theorem:tangent-vector-Legendre-submanifold-psi}.
\qed
%%%%%%%%
\end{Proof}
%%%%%%%%%
%%%%%%%
\begin{Remark}
%%%%%%%
It follows that 
$$
\Phi_{\cC\cA\varphi}^{\ \ \ \ *}\left[\,\lambda(Y_{\varphi}^{i})\,\right]
=0
\quad
\mbox{and}\quad
\lambda\left(\left.Y_{\varphi}^i\right|_{\Phi_{\cC\cA\varphi}\cA_{\varphi}}\right)
=0,\quad
\qquad i\in\{1,\ldots,n\}.
$$
%%%%%%%
\end{Remark}
%%%%%%%

%%%%%%%%%%%%%%%%%%%%%%%%%%%%%%%%%
\subsection{Calculations of inner product of quasi-static process and 
relaxation process  }
%%%%%%%%%%%%%%%%%%%%%%%%%%%%%%%%%%%%%%%%%%5
Given a metric tensor field on a contact manifold,  
one is able to calculate inner products of given two vector fields.  
In this subsection, such inner products are calculated 
for various vector fields including 
$\{Y_j^{\psi}\}$ given in \fr{Legendre-submanifold-tangent-vector-psi}  
and $X_h^{\psi}$ given in \fr{contact-Hamiltonian-Lyapunov-psi-vector}.
In addition, inner products are calculated for vector fields including 
$\{Y_{\varphi}^i\}$ given in \fr{Legendre-submanifold-tangent-vector-varphi}  
and $X_h^{\varphi}$ given in \fr{contact-Hamiltonian-Lyapunov-varphi-vector}. 

To give geometric characterization of the introduced contact Hamiltonian
system, one introduces the normalized vector fields follows.
%%%%%%%%%%
\begin{Def}
%%%%%%%%   
(Unit normalized vector field of contact Hamiltonian vector field $1$) : 
Let 
$X_h^{\psi}$ be the contact Hamiltonian vector field stated in
Theorem \ref{theorem:main-atrractor-psi-1},
and
$G$ 
the Mrugala metric tensor field \fr{def-Mrugala-metric}.
The vector field
\beq
U_h^{\psi}
:=\frac{1}{\|X_h^{\psi}\|}X_h^{\psi},\qquad
\|X_h^{\psi}\|
:=\sqrt{G(X_h^{\psi},X_h^{\psi})}\,,
\label{def-normalized-Hamiltonian-vector-psi}
\eeq
is referred to as the normalized vector field of $X_h^{\psi}$. 
%%%%%%%%%%
\end{Def}
%%%%%%%%   
%%%%%%%%%%
\begin{Remark}
%%%%%%%%%%\
Observe that
$G(U_h^{\psi},U_h^{\psi})=1$, and note that 
$U_{h}^{\psi}$ is not a contact Hamiltonian vector field whose contact 
Hamiltonian is $h_{\psi}$.
%%%%%%%%%%
\end{Remark}
%%%%%%%%%%\

%%%%%%%%%%%
\begin{Thm}
%%%%%%%%
\label{theorem:characterization-relaxation-process-psi}
(Characterization of relaxation processes $1$) : 
Let 
$R$ be the Reeb vector field, $\psi\in\GamLamC{0}$ a 
strictly convex function of $x$ only,    
$h_{\psi}$ and $X_h^{\psi}$ 
a contact Hamiltonian and its contact Hamiltonian vector field  
stated in 
Theorem \ref{theorem:main-atrractor-psi-1}, 
$U_{h}^{\psi}$ the normalized vector field defined in 
\fr{def-normalized-Hamiltonian-vector-psi},
$G$  
the Mrugala metric tensor field \fr{def-Mrugala-metric}, and  
$\nabla$ the Levi-Civita connection uniquely 
determined by $G$.
Furthermore let  
$\{Y_j^{\psi}\}$ be the basis given by 
\fr{Legendre-submanifold-tangent-vector-psi}.
Then, it follows that 
$$
G(Y_j^{\psi},R)
=0,\qquad
G(X_h^{\psi},R)
=h_{\psi},\qquad
G(X_h^{\psi},X_{h}^{\psi})
=(h_{\psi})^2.
$$
In addition, one has
$$
\nabla_{U_h^{\psi}}R^{\sharp}
=-\,\frac{1}{2}\left(
\frac{\partial\psi}{\partial x^j}-p_j
\right)\frac{1}{\wh{h}}
\frac{\dr \wh{h}}{\dr\Delta_{\psi}}\dr x^a,
\qquad
G(U_h^{\psi},Y_a^{\psi})
=\left(\,\frac{\partial\psi}{\partial x^a}-p_a\,\right)
\left(1+\frac{1}{2\,\wh{h}}\frac{\dr\wh{h}}{\dr\Delta_{\psi}}\right),
$$
and
$$
G(Y_a^{\psi},Y_b^{\psi})
=\frac{\partial^2\psi}{\partial x^a\partial x^b}
+\left(\frac{\partial\psi}{\partial x^a}-p_a\right)
\left(\frac{\partial\psi}{\partial x^b}-p_b\right).
$$

On $\Phi_{\cC\cA\psi}\,\cA_{\psi}$, one has
$$
\left.G(X_h^{\psi},X_h^{\psi})\right|_{\Phi_{\cC\cA\psi}\cA_{\psi}}
=0,\qquad
\left.G(Y_a^{\psi},Y_b^{\psi})\right|_{\Phi_{\cC\cA\psi}\cA_{\psi}}
=\frac{\partial^2\psi}{\partial x^a\partial x^b}.
$$
%%%%%%%%%%%
\end{Thm}
%%%%%%%%
%%%%%%%%
\begin{Proof}
%%%%%%%
These relations are verified by straightforward calculations.
From the expression of $X_h^{\psi}$ given by  
\fr{contact-Hamiltonian-Lyapunov-psi-vector},
that of $Y_j^{\psi}$ given by 
\fr{Legendre-submanifold-tangent-vector-psi},
 and that of $\lambda$, one has 
$$
\dr p_k(X_h^{\psi})
=\dot{p}_k,\qquad
\dr x^k(X_h^{\psi})
=0,\qquad
\lambda(X_h^{\psi})
=h_{\psi},
$$
$$
\dr p_k(Y_a^{\psi})
=\frac{\partial^2\psi}{\partial x^a\partial x^k},\qquad
\dr x^k(Y_j^{\psi})
=\delta_j^k,\qquad
\lambda(Y_j^{\psi})
=\frac{\partial\psi}{\partial x^j}-p_j.
$$
These equations and \fr{Reeb-vector-metric-dual-normalize}
are to be used in the following calculations. 
Substituting these equations into $G$ and $Y_a^{\psi}$, one has
$$
G(X_h^{\psi},-)
=\frac{1}{2}\dot{p}_j\dr x^j+h_{\psi}\,\lambda,\qquad
G(Y_a^{\psi},-)
=\frac{1}{2}\dr p_a+\frac{1}{2}\frac{\partial^2\psi}{\partial x^a\partial x^k}
\dr x^k+\left(\frac{\partial\psi}{\partial x^a}-p_a\right)\lambda,
$$
from which 
$$
\|X_h^{\psi}\|^2
=G(X_h^{\psi},X_h^{\psi})
=(h_{\psi})^2,\quad
G(Y_j^{\psi},R)
=0,
\qquad
G(X_h^{\psi},R)
=h_{\psi},
$$
and the explicit form of $U_h^{\psi}$ as 
$$
U_h^{\psi}
=\frac{1}{\wh{h}}X_h^{\psi}
=\frac{1}{\wh{h}}\left(\dot{p}_j\frac{\partial}{\partial p_j}
+\dot{z}\frac{\partial}{\partial z}\right)
=\left(\,\frac{\partial\psi}{\partial x^j}-p_j\,\right)
\,\frac{1}{\wh{h}}\,\frac{\dr \wh{h}}{\dr \Delta_{\psi}}
\frac{\partial}{\partial p_j}
+\frac{\partial}{\partial z}.
$$

To calculate $\nabla_{U_h^{\psi}}R^{\sharp}\in\GamLamC{1}$ with $R$ being Killing  
$\cL_RG=0$, one uses the formula 
\beq
\nabla_ZK^{\sharp}
=\frac{1}{2}\ii_Z\dr K^{\sharp},
\label{killing-formula-connection}
\eeq
where $K$ is a Killing vector field $\cL_KG=0$, and $Z$ arbitrary 
vector field. A proof of this formula is given in the Appendix B. 
Applying this formula with $K=R$, $K^{\sharp}=R^{\sharp}=\lambda$, and
$Z=U_h^{\psi}$, one has
$$
\nabla_{U_h^{\psi}}R^{\sharp}
=\frac{1}{2}\ii_{U_h^{\psi}}\dr\lambda
=\frac{1}{2}\ii_{U_h^{\psi}}\left(-\,\dr p_j\wedge\dr x^j\right)
=-\,\frac{1}{2}\frac{\dot{p}_j}{\wh{h}}\dr x^j
=-\,\frac{1}{2}\left(
\frac{\partial\psi}{\partial x^j}-p_j
\right)\frac{1}{\wh{h}}\frac{\dr \wh{h}}{\dr\Delta_{\psi}}\dr x^j.
$$
One calculates  
$$
G(U_h^{\psi},Y_a^{\psi})
=\frac{1}{2}\frac{\dot{p}_j}{\wh{h}}\dr x^j(Y_a)+\lambda(Y_a^{\psi})
=\left(\,\frac{\partial\psi}{\partial x^a}-p_a\,\right)
\left(1+\frac{1}{2\,\wh{h}}\frac{\dr\wh{h}}{\dr\Delta_{\psi}}\right)
$$
and
$$
G(Y_a^{\psi},Y_b^{\psi})
=\frac{\partial^2\psi}{\partial x^a\partial x^b}
+\left(\frac{\partial\psi}{\partial x^a}-p_a\right)
\left(\frac{\partial\psi}{\partial x^b}-p_b\right).
$$

To verify the relations on $\Phi_{\cC\cA\psi}\cA_{\psi}$, one uses 
$$
\frac{\partial\psi}{\partial x^j}-p_j
=0,\qquad j\in\{1,\ldots,n\},
$$
on $\Phi_{\cC\cA\psi}\,\cA_{\psi}$ due to  \fr{example-psi-Legendre-submanifold}.
It then follows that 
$$
G(X_h^{\psi},X_h^{\psi})|_{\Phi_{\cC\cA\psi}\,\cA_{\psi}}
=0,\qquad
\left.G(Y_a^{\psi},Y_b^{\psi})\right|_{\Phi_{\cC\cA\psi}\cA_{\psi}}
=\frac{\partial^2\psi}{\partial x^a\partial x^b}.
$$
\qed
%%%%%%%%
\end{Proof}
%%%%%%%
%%%%%%%%%%
 \begin{Remark}
% %%%%%%%%%
It has been shown from 
\fr{Reeb-vector-metric-geodesic} 
that the Reeb vector field $R$ gives geodesics and 
it can be shown that integral curves of $X_h^{\psi}$ are not geodesics. 
%%%%%%%%%%%
 \end{Remark}
%%%%%%%%%
%%%%%%%%%%%
\begin{Remark}
%%%%%%%%%
It follows from the conditions for $h_{\psi}$ on $\cD_{\psi}$, one has 
the expansion 
$\wh{h}(\Delta_{\psi})=\gamma_1\Delta_{\psi} +\gamma_2\Delta_{\psi}^2+\cdots$ 
with $\gamma_1>0$ and some $\gamma_2$. Thus, the term  
$$
\frac{1}{\wh{h}}\frac{\dr \wh{h}}{\dr \Delta_{\psi}}
=\frac{1+2\gamma_2\Delta_{\psi} +\cdots}{\gamma_1\Delta_{\psi} 
  +\gamma_2\Delta_{\psi}^2+\cdots}
$$
is divergent on the attractor where it follows that $\Delta_{\psi}=0$.
%%%%%%%%%%%
\end{Remark}
%%%%%%%%%

% %%%%%%%%%%
\begin{Remark}
%%%%%%%%%
The norm $\|X_h^{\psi}\|$ becomes smaller as approaching 
to the Legendre submanifold $\Phi_{\cC\cA\psi}\,\cA_{\psi}$. 
Thus, to discuss geometry involving  
the contact Hamiltonian system around $\Phi_{\cC\cA\psi}\,\cA_{\psi}$ 
 an appropriate vector field 
is $U_{h}^{\psi}$, rather than $X_h^{\psi}$.  
% %%%%%%%%%%
 \end{Remark}
%%%%%%%%%

In addition, one has the following.
%%%%%%%%%%
\begin{Def}
%%%%%%%%   
(Unit normalized vector field of contact Hamiltonian vector field $2$) : 
Let 
$X_h^{\varphi}$ be the contact Hamiltonian vector field stated in
Theorem \ref{theorem:main-atrractor-varphi-1}, and
$G$  
the Mrugala metric tensor field \fr{def-Mrugala-metric}.
The vector field 
\beq
U_h^{\varphi}
:=\frac{1}{\|X_h^{\varphi}\|}X_h^{\varphi},\qquad
\|X_h^{\varphi}\|
:=\sqrt{G(X_h^{\varphi},X_h^{\varphi})},
\label{def-normalized-Hamiltonian-vector-varphi}
\eeq 
is referred to as the normalized vector field of $X_h^{\varphi}$.
%%%%%%%%%%
\end{Def}
%%%%%%%%   
%%%%%%%%%%
\begin{Remark}
%%%%%%%%%%\
Observe that 
$G(U_h^{\varphi},U_h^{\varphi})=1,$ and note that 
$U_{h}^{\psi}$ is not a contact Hamiltonian vector field whose contact 
Hamiltonian is 
$h_{\varphi}$.
%%%%%%%%%%
\end{Remark}
%%%%%%%%%%\

%%%%%%%%%%%
\begin{Thm}
%%%%%%%%
(Characterization of relaxation processes $2$) : 
Let 
$R$ be the Reeb vector field, 
$\varphi\in\GamLamC{0}$ a strictly 
convex function 
of $p$ only,    
 $h_{\varphi}$ and $X_h^{\varphi}$ 
 a contact Hamiltonian and its contact Hamiltonian vector field 
stated in 
Theorem \ref{theorem:main-atrractor-varphi-1}, 
$U_h^{\varphi}$  the normalized vector field defined in
\fr{def-normalized-Hamiltonian-vector-varphi}, 
$G$  
the Mrugala metric tensor field \fr{def-Mrugala-metric}, 
and $\nabla$ the Levi-Civita connection uniquely 
determined by $G$. 
Furthermore let  
$\{Y_{\varphi}^i\}$ be the basis given by 
\fr{Legendre-submanifold-tangent-vector-varphi}.
Then, it follows that 
$$
G(Y_{\varphi}^a,R)
=0,\qquad
G(X_h^{\varphi},R)
=h_{\varphi},\qquad
G(X_h^{\varphi},X_{h}^{\varphi})
=(h_{\varphi})^2.
$$
In addition, one has
$$
\nabla_{U_h^{\varphi}}R^{\sharp}
=\frac{1}{2}\left(
\frac{\partial\varphi}{\partial p_i}-x^i
\right)\frac{1}{\wh{h}}\frac{\dr \wh{h}}{\dr\Delta_{\varphi}}\dr p_i,
\qquad
G(U_h^{\varphi},Y_{\varphi}^i)
=\frac{1}{2}\left(\frac{\partial\varphi}{\partial p_i}-x^i\right)
\frac{1}{\wh{h}}
\frac{\dr \wh{h}}{\dr\Delta_{\varphi}}.
$$ 
On $\Phi_{\cC\cA\varphi}\,\cA_{\varphi}$, one has
$$
\left.G(X_h^{\varphi},X_h^{\varphi})\right|_{\Phi_{\cC\cA\varphi}\,
\cA_{\varphi}}
=0,
\qquad 
\left.G(Y_{\varphi}^a,Y_{\varphi}^b)\right|_{\Phi_{\cC\cA\varphi}\,\cA_{\varphi}}
=\frac{\partial^2\varphi}{\partial p_a\partial p_b}.
$$
%%%%%%%%%%%
\end{Thm}
%%%%%%%%
%%%%%%%%
\begin{Proof}
%%%%%%%
A way to prove this is analogous to 
the proof of Theorem\,\ref{theorem:characterization-relaxation-process-psi}. 
\qed
%%%%%%%%%%%
\end{Proof}
%%%%%%%%%%%

%%%%%%%%%%
 \begin{Remark}
% %%%%%%%%%
It has been shown from 
\fr{Reeb-vector-metric-geodesic} 
that the Reeb vector field $R$ gives geodesics and 
it can be shown that integral curves of $X_h^{\varphi}$ are not geodesics. 
%%%%%%%%%%%
 \end{Remark}
%%%%%%%%%
%%%%%%%%%%%
\begin{Remark}
%%%%%%%%%
It follows from the conditions for $h_{\varphi}$ on $\cD_{\varphi}$, one has 
the expansion 
$\wh{h}(\Delta_{\varphi})=\gamma_1\Delta_{\varphi} +\gamma_2\Delta_{\varphi}^2+\cdots$ 
with $\gamma_1>0$ and some $\gamma_2$. Thus, the term  
$$
\frac{1}{\wh{h}}\frac{\dr \wh{h}}{\dr \Delta_{\varphi}}
=\frac{1+2\gamma_2\Delta_{\varphi} +\cdots}{\gamma_1\Delta_{\varphi} +\gamma_2\Delta_{\varphi}^2+\cdots}
$$
is divergent on the attractor where it follows that $\Delta_{\varphi}=0$.
%%%%%%%%%%%
\end{Remark}
%%%%%%%%%

% %%%%%%%%%%
\begin{Remark}
%%%%%%%%%
The norm $\|X_h^{\varphi}\|$ becomes smaller as approaching 
to the Legendre submanifold $\Phi_{\cC\cA\varphi}\,\cA_{\varphi}$.  
Thus, to discuss geometry involving 
the contact Hamiltonian system around 
$\Phi_{\cC\cA\varphi}\,\cA_{\varphi}$ 
an appropriate vector field 
is $U_{h}^{\varphi}$, rather than $X_h^{\varphi}$.  
% %%%%%%%%%%
 \end{Remark}
%%%%%%%%%

%%%%%%%%%%
%%%%%%%%%%%%%%%%%%%%%%%%%%%%%%%%%%%%%%%%%%%5
 \section{Concluding remarks}
\label{sec-summary}
%%%%%%%%%%%%%%%%%%%%%%%%%%%%%%%%%%%%%%%%
This paper offers a view point that a class of 
relaxation processes can be treated as contact Hamiltonian vector fields  
on a contact manifold,  
by postulating that Legendre submanifolds are physically interpreted as 
equilibrium states. 
These vector fields on a contact manifold have been characterized with 
a metric tensor field, 
and relations between the contact Hamiltonian and lower-dimensional  
spaces of a contact manifold have been clarified.  
In addition,  it has been shown 
that a contact manifold and a strictly convex function induce 
a dually flat space.
Thus, this paper provides a view point 
that ideas in contact geometry can be used to study information geometry 
in addition to thermodynamics.
Throughout this paper, Legendre duality  has explicitly been stated 
outside the Legendre submanifold where a Legendre submanifold is given. 
We feel that this is important since 
Legendre duality is usually discussed at equilibrium states, 
and how it is important at nonequilibrium states should be clarified.   

There are numbers of extensions that follow from this work. 
One of such a future work is 
to give physical meanings of claimed theorems. 
More precisely, if 
a contact Hamiltonian is derived from a microscopic dynamical 
model, then the meaning of such a contact Hamiltonian becomes clear.  
In connection with other forms of the 
geometrization of nonequilibrium statistical mechanics, 
it is interesting to see a relation between 
this work and the one in Ref.\cite{Mrugala1992}. 
In addition to these, it is important to elucidate a link between our 
methodology and that of Ref.\cite{OW09}  
in which 
a relaxation process of a nonlinear diffusion equation was analyzed 
by introducing a statistical manifold.
Although it is 
expected that the higher dimensional manifold used in Ref.\cite{OW09} 
can express much wider class of nonequilibrium processes 
than that of this paper,  
the limitations of the both approaches for expressing nonequilibrium 
processes are not known. 
We believe 
that the elucidation of these remaining questions 
will develop the theory of geometric nonequilibrium thermodynamics.

%%%%%%%%%%%
\section*{Acknowledgments }
%%%%%%%%%%%
The author would like to thank Y. Shikano and 
Institute for Molecular Science for supporting my work, 
and thank M. Koga, T. Wada, and an anonymous referee 
for giving various comments on this paper. 
 
%%%%%%%%%%%
\appendix 
%%%%%%%%%%%%
%%%%%%%%%%%%%%%%%%%%%%%%%55
\section{Appendix : Derivation of the kinetic spin model without 
spin-coupling}
%%%%%%%%%%%%%%%%%%%%%
In this appendix, the dynamical system 
\fr{kinetic-spin-model}, 
$$
\frac{\dr \ave{\sigma}}{\dr t}
=\gamma^{\,\prime}\left(\,\tanh(\beta_{\abs}H)-\ave{\sigma}\,\right),
$$
is derived. Here, $\gamma^{\,\prime}$ is a constant,
and $\ave{\sigma}$ a function of $t$.

Consider the master equation, 
$$
\frac{\dr}{\dr t}\mbbP_{\theta}(\sigma,t)
=-W(\sigma\mapsto-\,\sigma)\,\mbbP_{\theta}(\sigma,t)
+W(-\,\sigma\mapsto\,\sigma)\,\mbbP_{\theta}(-\,\sigma,t),
$$
where $\mbbP_{\theta}(\sigma,t)$ is a probability that 
a state $\sigma$ is realized at time $t$, $W(\sigma\mapsto-\,\sigma)$  the 
transition rate from a state $\sigma$ to $-\,\sigma$ for a unit time. 

The explicit form of $W(\sigma\mapsto-\,\sigma)$ is determined as follows.
Demanding the detailed balance condition 
$$
W(\sigma\to -\,\sigma)\,\mbbP_{\theta}^{\,\can}(\sigma)
=W(-\,\sigma\to \sigma)\,\mbbP_{\theta}^{\,\can}(-\sigma),
$$
and using the explicit form of $\mbbP_{\theta}^{\,\can}(\sigma)$ given by 
\fr{canonical-distribution-Ising-model} with 
\fr{partition-function-canonical-distribution-Ising-model}, 
one has
$$
\frac{W(\sigma\mapsto\,-\,\sigma)}{W(-\,\sigma\mapsto\,\sigma)}
=\frac{\exp(-\theta\,\sigma)}{\exp(\theta\,\sigma)}
=\exp(-2\,\theta\,\sigma)
=\frac{1-\sigma\tanh\theta}{1+\sigma\tanh\theta}.
$$
A solution to this equation for $W$ is found to be
$$
W(\sigma\mapsto-\,\sigma)
=\frac{\gamma^{\,\prime}}{2}\left(\,
1-\sigma\tanh\theta
\,\right),
$$
where $\gamma^{\,\prime}$ is a constant. 

To obtain an equation for an averaged quantity, one defines the average of 
arbitrary function of $\sigma$ as 
$$
\ave{f}(t)
:=\sum_{\sigma=\pm1}f(\sigma)\,\mbbP_{\theta}\,(\sigma,t).
$$
Choosing $f$ as $\sigma$ and differentiating $\ave{\sigma}(t)$ 
with respect to $t$, one has 
$$
\frac{\dr}{\dr t}\ave{\sigma}(t)
=\sum_{\sigma}\sigma\,\frac{\dr}{\dr t}\mbbP_{\theta}(\sigma,t)
=\sum_{\sigma}\sigma\,\left[\,
-W(\sigma\mapsto-\,\sigma)\,\mbbP_{\theta}(\sigma,t)
+W(-\,\sigma\mapsto\,\sigma)\,\mbbP_{\theta}(-\,\sigma,t)
\,\right],
$$ 
and then substituting the explicit form of $W$, one has
$$
\frac{\dr}{\dr t}\ave{\sigma}(t)
=\gamma^{\,\prime}\left(\,-\,\ave{\sigma}(t)+\,\tanh\theta\,\right),
$$
which is the same as \fr{kinetic-spin-model}. 
%%%%%%%%%%%%%%%%%%%%%%%%%55
\section{Appendix : Derivation of formulas used in \S\ref{sec-calculations-with-metric} }
%%%%%%%%%%%%%%%%%%%%%
In this appendix, the formulas used in 
\S\ref{sec-calculations-with-metric} are proved. 

Let $(\cM,g)$ be an $n$-dimensional Riemannian or 
a pseudo-Riemannian manifold,
$\{e^a\}$ the $g$-orthonormal co-frame such that 
$$
g=\eta_{ab}\,e^a\otimes e^b,
$$
with $\{\eta_{ab}\}=\diag{\pm 1,\cdots,\pm1}$,
$\{X_a\}$ dual of $\{e_a\}$ satisfying $e^a(X_b)=\delta_b^a$, $K$ a 
Killing vector field $\cL_Kg=0$, 
$\star : \GamLamM{q}\to\GamLamM{n-q},(q\in\{0,\ldots,n\})$ the 
Hodge dual map with $\star 1$ being the canonical volume form 
$\star 1:=e^1\wedge\cdots\wedge e^n$, $\star^{-1}$ the inverse map of $\star$,  
$\nabla$  the Levi-Civita connection, and  
$\delta:\GamLamM{q}\to\GamLamM{q-1}$  the co-derivative defined to be 
\beq
\delta\,\alpha
=\star^{-1}\dr\star  (-)^q\alpha,
\label{definition-delta-co-derivative}
\eeq
for arbitrary $\alpha\in\GamLamM{q}$. 
Throughout this appendix, the metric dual of a vector field $Z$ 
is denoted $Z^{\sharp}:=g(Z,-)$.
%%%%%%%%%%%
\subsection{Derivation of the formula\, \fr{Laplacian-and-Killing-formula}}
%%%%%%%%%%

In what follows, formula \fr{Laplacian-and-Killing-formula},    
$$
\star^{-1}\dr\star\dr f
=0,
$$
with $f\in\GamLamM{0}$ being such that $\dr f=g(K,-)$, 
is proved. 
%%%%%%%%5
\begin{Proof}
%%%%%%%%%
It follows for a $q$-form $\alpha$ that 
$$
\delta\,\alpha
=-\,\eta^{ab}\ii_{X_b}\nabla_{X_a}\alpha,
$$
where $\eta^{ab}$ is such that $\eta_{ac}\eta^{cb}=\delta_{a}^{b}$.
Combining this equation and 
\fr{definition-delta-co-derivative}, one has for a one-form $\alpha$ 
$$
\star^{-1}\dr\star  (-)^1\alpha
=-\,\eta^{ab}\ii_{X_b}\nabla_{X_a}\alpha.
$$
Substituting $\alpha=g(K,-)=:K^{\sharp}\in\GamLamM{1}$ into the above equation,  
one has
$$
\star^{-1}\dr\star  K^{\sharp}
=\eta^{ab}\ii_{X_b}\nabla_{X_a} K^{\sharp}.
$$
The right hand side can be written as
$$
\eta^{ab}\ii_{X_b}\nabla_{X_a} K^{\sharp}
=\eta^{ab}\ii_{X_b}(\,\nabla_{X_a} K\,)^{\sharp}
=\eta^{ab}g(\,\nabla_{X_a} K\,,X_b)
=-\,\eta^{ab}g(\,\nabla_{X_b} K\,,X_a)
=-\,\eta^{ab}\ii_{X_b}\nabla_{X_a} K^{\sharp},
$$
where we have used $\eta^{ab}=\eta^{ba}$ and the Killing equation
$$
g(\nabla_YK,Z)
=-\,g(\nabla_ZK,Y),
$$
for arbitrary $Y,Z\in\GTM$. 
Thus 
$$
\eta^{ab}\ii_{X_b}\nabla_{X_a} K^{\sharp}
=0,
$$
from which 
$$
\star^{-1}\dr\star  K^{\sharp}
=0.
$$
In the special case where $K^{\sharp}=\dr f$ with $f\in\GamLamM{0}$, 
one immediately has 
$$
\star^{-1}\dr\star  \dr f
=0.
$$
\qed
%%%%%%%%5
\end{Proof}
%%%%%%%%%
%%%%%%%%%%%%%%%%%%%%%%%%%%%%%%%%%%%%%%%
\subsection{Derivation of the formula\,\fr{killing-formula-connection}
}
%%%%%%%%%%
In what follows, \fr{killing-formula-connection},
$$
\nabla_ZK^{\sharp}
=\frac{1}{2}\ii_Z\dr K^{\sharp}, 
$$
with $Z\in\GTM$ being an arbitrary vector field, is proved.
%%%%%%%%5
\begin{Proof}
%%%%%%%%%
Decompose $Z$ in terms of the basis $\{X_a\}$ as 
$$
Z=Z^aX_a,
$$
where $\{Z^a\}$ is a set of functions. It follows for arbitrary vector fields  
$Y$ and $Z$ that 
$$
Z^a
=\ii_Z e^a,\qquad
\nabla_Z
=Z^a\nabla_{X_a},\qquad
\dr Y^{\sharp}
=e^a\wedge\nabla_{X_a}Y^{\sharp},\qquad
e^a\wedge\left(\ii_{X_a}\nabla_Z Y^{\sharp}\right)
=\nabla_Z Y^{\sharp}.
$$ 
From these equations, one has
\beqa
\ii_Z\dr K^{\sharp}
&=&\left(\ii_{Z}e^a\right)\nabla_{X_a}K^{\sharp}
-e^a\,\left(\ii_Z\nabla_{X_a}K^{\sharp}\right)
=\nabla_{Z}K^{\sharp}-e^a\,\left(\ii_Z\nabla_{X_a}K^{\sharp}\right)
\non\\
&=&\nabla_ZK^{\sharp}-e^a\wedge \left(\ii_{Z}\nabla_{X_a}K^{\sharp}\right)
\non\\
&=&\nabla_ZK^{\sharp}-e^a\wedge \left(\ii_{Z}\nabla_{X_a}K^{\sharp}\right)
+\left\{
-e^a\wedge \left(\ii_{X_a}\nabla_ZK^{\sharp}\right)
+e^a\wedge \left(\ii_{X_a}\nabla_ZK^{\sharp}\right)
\right\}
\non\\
&=&
\nabla_ZK^{\sharp}-e^a\wedge \left(
\ii_{Z}\nabla_{X_a}K^{\sharp}
+\ii_{X_a}\nabla_ZK^{\sharp}\right)+\nabla_ZK^{\sharp}.
\non
\eeqa
Applying the Killing equation 
$$
\ii_{Z}\nabla_{X_a}K^{\sharp}
+\ii_{X_a}\nabla_ZK^{\sharp}
=0,
$$
one has
$$
\ii_Z\dr K^{\sharp}
=2\nabla_ZK^{\sharp}.
$$
\qed
%%%%%%%%%%%%%
\end{Proof}
%%%%%%%%%%%%%

%%%%%%%%%%%%%%%%%%%%%%%%%%%

%%%%%%%%%%%%%%%%%%%%%%
%%%%%%%%%%%%%%%
\end{document}